\documentclass[aps,superscriptaddress,showpacs,nofootinbib,eqsecnum]{revtex4}

\usepackage{amsmath}
\usepackage{mathrsfs}
\usepackage{multirow}
\usepackage{hhline}
\usepackage{epsfig} 
\usepackage{color} 
\usepackage{hyperref}
\hypersetup{
     colorlinks   = true,
     linkcolor    = black,
     urlcolor     = magenta,
     citecolor    = blue
}

\input{epsf}
\def\be{\begin{equation}} \def\ee{\end{equation}} \def\bea{\begin{eqnarray}}
\def\eea{\end{eqnarray}}

\def\beq{\begin{equation}}
\def\eeq{\end{equation}}
\def\beqa{\begin{eqnarray}}
\def\eeqa{\end{eqnarray}}

\begin{document}

\title{Coexistence of Multiple Phases in Magnetized Quark Matter with  Vector Repulsion}

\author{Robson Z. Denke} \email{r.denke@posgrad.ufsc.br}
\affiliation{Departamento de F\'{\i}sica, Universidade Federal de Santa
  Catarina, 88040-900 Florian\'{o}polis, Santa Catarina, Brazil}
  
\affiliation{Departamento de F\'{\i}sica, Funda\c c\~ao Universidade Regional de Blumenau, 89012-900 Blumenau, Santa Catarina, Brazil}

\author{Marcus Benghi Pinto} \email{marcus.benghi@ufsc.br}
\affiliation{Departamento de F\'{\i}sica, Universidade Federal de Santa
  Catarina, 88040-900 Florian\'{o}polis, Santa Catarina, Brazil}

\begin{abstract}
We   explore the phase structure of dense magnetized quark matter when  a repulsive vector interaction, parametrized by $G_V$, is present.  Our  results show that for a given  magnetic field intensity ($B$) one may find a value of $G_V$ for which quark matter may coexist at three different baryonic density values leading to the appearance of two triple points in the phase diagram which have not been observed before.  Another novel result is that at high pressure and low temperature we  observe a first order transition which presents a negative slope in the $P-T$ that is reminiscent of the   solid-liquid transition line  observed within the water phase diagram. These unusual patterns occur for $G_V$ and $B$ values which lie within the range presently considered in many  investigations related to the study of magnetars.
 
\end{abstract}


\maketitle

\section{Introduction}

Strongly interacting  magnetized matter may be produced in non central relativistic heavy ion collisions \cite {kharzeev}   and may also be present in magnetars \cite {magnetars} reaching up about to $\sim 10^{19} \,G$ and $\sim 10^{18} \,G$ in each of these two physical situations. As far as heavy ion-collisions are concerned the presence of a strong magnetic field
most certainly plays a role  despite the fact that,   in principle,
the   field intensity should decrease very rapidly  
lasting for about 1-2 fm/c only \cite {kharzeev}. The possibility that
this short time interval may    
\cite{tuchin} or may not \cite{mclerran} be affected by conductivity
remains under dispute.

A striking property expected to occur in such extreme conditions  is  the so called  Magnetic Catalysis (MC) phenomenon \cite{MC}, which implies that the order parameter for the chiral transition represented by the quark-antiquark condensate rises as the field becomes more intense 
(see Ref. \cite {reviewMC} for a review).  It is then natural to ask how the expected quantum chromodynamics (QCD)  transitions are affected by the presence of intense magnetic fields, a question which has been addressed in   many recent works  (see  Refs. \cite {reviews} for an updated discussion). To summarize the main results obtained in these investigations let us start by recalling that at  finite temperatures and vanishing chemical potential   both, 
  model approximations \cite {eduardo, prd, newandersen, prd12} and  lattice QCD (LQCD) evaluations \cite {earlylattice, lattice} 
  agree that a cross over, which is predicted to occur in the the absence of magnetic fields \cite {Aoki},  persists when strong magnetic fields are 
  present. However, a source of disagreement between recent LQCD evaluations 
  \cite {lattice} and model predictions regards the behavior of the pseudocritical temperature ($T_{\rm pc}$), at which the cross over 
  takes place, as a function of the magnetic field intensity. The LQCD simulations of Ref. \cite {lattice}, performed with $2+1$ 
  quark flavors and physical pion mass values, predict that $T_{\rm pc}$ should decrease with $B$ while early model evaluations 
  predict an increase (see Ref. \cite {prd} and references therein).  This problem has been recently addressed by different groups \cite{catalysis-ours} which basically agree that the different results stem from the fact that most effective models miss back reaction effects (the indirect interaction of gluons and $B$) as well as the QCD asymptotic freedom phenomenon.
On the other hand, LQCD results are currently unavailable at high densities and low temperatures so that one has to rely in model approximations \cite{prd,newandersen,prd12} which  predict  that  the first order  chiral transition takes place  at a  coexistence chemical potential value which is lower than the one observed  for the case of unmagnetized matter leading to the phenomenon  called  Inverse Magnetic Catalysis (IMC) \cite{IMC} (see Ref. \cite{preis} for a 
 physically intuitive discussion of the IMC 
phenomenon). Further investigations performed with the Nambu--Jona-Lasinio model (NJL)  have been carried out to   locate  the critical end
point  (CP) as well as the coexistent
chemical potentials associated with the first order chiral transition \cite{prd,prd12}. The results suggest that the size of first order 
transition line increases as the field becomes stronger affecting the
position of the (second order) CP where the first
order transition line terminates.  At the same time, the size and 
location of the coexistence region in the presence of a magnetic field appears to oscillate around the $B=0$ values \cite{Denke:2013gha}.  Together, 
 all these effects have interesting consequences
for quantities which depend on the details of the coexistence region such as the surface tension   and may have consequences for studies related to the properties of magnetized compact stellar objects \cite {andre}.
The motivation for  the present investigation  stems from an early work by Ebert and collaborators \cite {ebert} who recognized that  the MC phenomenon, associated to the
filling of Landau levels, could lead to more exotic phase
transition patterns as a consequence of the induced magnetic
oscillations. To confirm this assumption these authors have
considered a wide range of the (scalar) coupling values for  the two flavor
NJL model at vanishing temperatures and,
as expected,  have  observed unusual phase structures as a function
of the  chemical potential such as an infinite number of massless
chirally symmetric phases, a cascade of massive phases with broken
chiral invariance among other features.
More recently, this seminal study has  been extended by a more
systematic, and numerically accurate analysis with two and three flavors in Refs.
 \cite{Allen:2013lda} and \cite{ Grunfeld:2014qfa} respectively. The results confirm that for certain model parametrizations  one is able to observe 
more than one first order phase transition, which is signaled when the thermodynamical potential develops two degenerate  minima at different values
of the coexistence chemical potential  \footnote{This fact has also been recently observed to arise within another
effective four fermion theory described by the $2+1$ d Gross-Neveu model \cite{enois}.}.
It is important to remark that, in general, {\it weak} first order transitions can be easily missed in a numerical evaluation due to the fact that
the two degenerate minima appear almost at the same location being separated by a tiny potential barrier so that
their study requires extra care. Physically, this corresponds  to a situation where  two different (but almost identical)
densities coexist at the same chemical potential, temperature and pressure.  Here, one of our goals is to extend the investigation of these cascades of first order phase transitions, observed in Refs. \cite {ebert, Allen:2013lda, Grunfeld:2014qfa}, when {\it hot} magnetized quark matter is subject to the presence of a  repulsive vector channel parametrized by $G_V$. This type of interaction  provides a saturation mechanism similar to those found 
in effective nuclear models \cite {walecka} and is known to be important for an accurate description of quark matter at high baryonic densities \cite{volker}. As we have verified in a previous work \cite{Denke:2013gha}, the repulsive vector coupling modifies the magnetic effects 
mainly at lower temperatures and plays an opposite role compared to $B$ in the QCD phase diagram. As a matter of fact, the increase of the magnetic field shifts the first order transition  to lower values of the coexistence chemical potential while a nonzero vector repulsion  produces the reverse effect. This feature has also been recently verified within the three flavor case \cite{Menezes:2014aka} in the analysis of compact stellar objects. 
Here, we shall see that  the presence of a vector repulsion allows for further  interesting possibilities associated with the chiral first order transition due to the fact that  this term can stabilize intermediary density magnetic phases. Being carried out at finite temperatures the present investigation also allows for a more complete description  of the coexistence region and makes it possible to better understand the  physical nature of the associated phase transitions.
As a first novelty we show that within a cascade of the transitions the one which takes place at the highest pressure value seems to be  reminiscent of the ``solid-liquid" type of transition displayed by the phase diagram of water while the others are of the usual ``liquid-gas" type (commonly observed within QCD effective theories).  Motivated by the fact that the Lennard-Jones potential, which describes water, also has a repulsive part we have scanned over the $G_V$ values to search for any eventual triple point in the phase diagram of magnetized quark matter since this situation cannot be completely ruled out in the scenario considered here. The reason is that  the presence of a magnetic field induces the free energy to develop multiple minima while  the repulsive interaction  favors stability  so that eventually  three (instead of the usual two) minima could be globally degenerate leading to the coexistence of three phases. As we shall demonstrate the numerical results have confirmed our expectations so that for very particular, but yet realistic, parameter values the phase diagram  of strongly interacting magnetized quark matter may indeed display triple points. 

The work is organized as follows. In the next section we present the  free energy in the presence of 
 a repulsive vector channel for the magnetized NJL model within  the mean field approximation (MFA) framework.
In Sec. III we discuss how a cascade of the usual type of first order phase transitions, with two degenerate minima,  takes place at finite temperatures.  In the same section we  extend the analysis to the finite temperature domain in order to draw the phase diagrams in the $T-\mu$, $T-\rho_B$ and $P-T$ planes. Section IV is devoted to the study of  unusual first order phase transitions where the free energy develops more than two degenerate minima allowing for the existence of triple points.  Our conclusions and final remarks are presented in Sec. V.

\section{ NJL Magnetized Free Energy with a Repulsive Vector Interaction}

The  QCD  interaction between quarks can be effectively described  by the well-known NJL theory \cite{njl} which reproduces,
at lower energies, the main features of chiral symmetry breaking. In the usual two flavor version the same coupling constant, $G_S$, sets   the interaction strength in both the scalar  and pseudo-scalar channels. However, in finite density investigations  the model produces more realistic results if an
 additional repulsive vector channel, parametrized by  $G_V$, is introduced. 
In this case, the corresponding Lagrangian density \cite{volker,buballa} can be written as

\begin{equation}
\mathscr{L} = \bar{\psi}(i \displaystyle \gamma_{\mu}\partial^{\mu}  -m)\psi + G_S[(\bar{\psi}\psi)^2+(\bar{\psi}i\gamma_5 \vec{\tau} \psi)^2]-
G_V(\bar{\psi}\gamma^{\mu}\psi)^2,
\end{equation}
\noindent
where $m=m_u \simeq m_d$ is the bare quark mass. 
In order to derive the thermodynamical potential within the MFA the quadratic interaction terms appearing in the above Lagrangian are linearized
by the introduction of the mean fields expressed in terms of the scalar condensate, $\phi=\langle {\bar \psi}\psi \rangle$, and the quark number  density, $\rho=\langle \psi^+ \psi \rangle$
\begin{equation}
(\bar{\psi} \psi)^2 \simeq 2 \phi  {\bar \psi} \psi -  \phi^2 \,\,\,\,{\rm and}\,\,\,\, 
(\bar{\psi} \gamma^0\psi)^2 \simeq 2  \rho  \psi^+ \psi - \rho^2 \,\,,
\end{equation}
where quadratic terms in the fluctuations have been neglected while the pseudo-scalar term does not  contribute at this level.
Then, in the case of symmetric quark matter ($\mu=\mu_u=\mu_d$) the theory is described by 
\begin{equation}
\mathscr{L} = \bar{\psi}(i \displaystyle \gamma_{\mu}\partial^{\mu}-M +\tilde{\mu}\gamma^0)\psi- \displaystyle{\frac{(M-m)^2}{4G_S}}+{\frac{(\mu-\tilde{\mu})^2}{4G_V}} \,\,\,,
\end{equation}
where   the effective quark mass, $M$, and the  effective chemical potential, ${\tilde \mu}$, are determined  upon applying the corresponding
minimization conditions, $\delta \Omega/ \delta M=0$ and  $\delta \Omega/ \delta {\tilde \mu}=0$. 
Integrating over the fermionic fields   yields  the thermodynamical potential

\begin{equation}
\Omega=\frac{(M-m)^2}{4G_S}-\frac{(\mu-\tilde{\mu})^2}{4G_V}   +\frac{i}{2}{\rm tr} \int \frac{d^4p}{(2\pi)^4} \ln  [-p^2+M^2] \,\,.
\label{free}
\vspace{0.4 cm}
\end{equation}
\noindent  
One can then include the effects of a static magnetic field and a thermal bath to this dense quark matter system by applying the following replacements \cite {eduana}  to Eq. (\ref{free}):

\begin{equation}
p_0\rightarrow i(\omega_{\nu}-i\mu)\,\, , \nonumber
\end{equation}

\begin{equation}
p^2 \rightarrow p_z^2+(2n+1-s)|q_f|B \,\,\,\,\, , \,\,\mbox{with} \,\,\, s=\pm 1 \,\,\, , \,\, n=0,1,2...\,\,, \nonumber
\end{equation}

\begin{equation}
\int_{-\infty}^{+\infty}  \frac{d^4p}{(2\pi)^4}\rightarrow i\frac{T |q_f| B}{2\pi}\sum_{\nu=-\infty}^{\infty}\sum_{n=0}^{\infty}\int_{-\infty}^{+\infty} \frac{dp_z}{2\pi} \,\,,\nonumber
\end{equation}
\noindent 
where $\omega_\nu=(2\nu+1)\pi T$ ($\nu=0,\pm1,\pm2...$) represents the Matsubara frequencies  for fermions. The Landau levels are labelled  by $n$ while the
absolute values of quark electric charges $|q_f|$ are $|q_u|= 2e/3$ and  $|q_d| = e/3$ with $e = 1/\sqrt{137}$ representing 
the electron charge
\footnote {Our results are expressed in Gaussian natural units where $1 \, {\rm MeV}^2 = 1.44 \times 10^{13}\, G$.}. 
Then, following Ref. \cite {prcsu2} 
we can  write the  thermodynamical potential as

\begin{equation}
\Omega ={\frac{(M-m)^2}{4G_S}}-{\frac{(\mu-\tilde{\mu})^2}{4G_V}}+{\Omega}_{vac}+\Omega_{mag}+{\Omega}_{med}\,\,\,,
\label{omega}
\end{equation}
where the vacuum contribution reads

\begin{equation}
{\Omega}_{vac} = -2N_cN_f\displaystyle\int {\frac{d^3{\bf p}}{(2\pi)^3}} \sqrt{p^2+M^2}\;\;,
\label{vac}
\end{equation}
and, as usual,  can be regularized by a non-covariant sharp cut-off, $\Lambda$, yielding

\begin{equation}
{\Omega}_{vac} = {\frac{N_cN_f}{8\pi^2}}  \left\{ M^4\ln{\left [
{\frac{(\Lambda+\epsilon_{\Lambda})}{M}}\right ]-\epsilon_{\Lambda}\Lambda[\Lambda^2+{\epsilon_{\Lambda}}^2
]}\right\} \,\,\,,
\end{equation}
where $\epsilon_{\Lambda}$ represents the energy $\sqrt{{\Lambda}^2+M^2}$ at the cutoff momentum value $\Lambda$. We remark that in Refs. \cite{noronha07,
  fukushima08} the authors choose a smooth cut off to avoid 
unphysical oscillations, which appear when the pairing interaction is
included because the sharp cut off limits the allowed momenta. 
In the present work, no superconducting phase (that would require
the pairing gap equation to be solved) is used and hence we do not
face the problem of unphysical solutions.
The magnetic contribution of the thermodynamical potential is given by 

\begin{equation}
{\Omega}_{mag}=-\displaystyle\sum_{f=u}^d{\frac{N_c(|q_f|B)^2}{2\pi^2}}
\biggr\{  \zeta'[-1,x_f]-{\frac{1}{2}}(x_f^2-x_f)\ln{x_f}+{\frac{x_f^2}{4}}\biggl \}\,,
\end{equation}
whre  we have used the definition $x_f= M^2/(2 |q_f| B)$ and the derivative of the Riemann-Hurwitz zeta function $\zeta'(-1, x_f) = d\zeta(z, x_f)/dz|_{z=-1}$ (see the appendix of Ref. \cite{prcsu2} for detailed steps). 
Finally, the  term ${\Omega}_{med}$ represents the in-medium contribution 

\begin{equation}
{\Omega}_{med} = -\displaystyle{\frac{N_c}{2\pi}} \sum_{f=u}^d \sum_{k=0}^{\infty}   \alpha_k(|q_f|B)\int_{-\infty}^{\infty} {\frac{dp_z}{2\pi}}
 \left\{T\ln{[1+e^{-(E_{p}+\tilde{\mu})/T}]}
+T\ln{[1+e^{-(E_{p}-\tilde{\mu})/T}]}\right\}\,\,\,,
\label{Omega med}
\end{equation}
where $\alpha_k = 2 - \delta_{k0}$ and $E_{p} = \sqrt{p_z^2+2k|q_f|B+M^2}$. A similar expression for the magnetized thermodynamical potential at $G_V=0$ was  originally obtained in  Ref.  \cite {klimenko} where  Schwinger's proper time approach has been used.
Solving $\delta \Omega/\delta M = 0$ and  $\delta \Omega/\delta \tilde{\mu} = 0$ we get the following coupled self consistent equations
\begin{equation}
 M = m-2G_S\phi\,\,, 
 \label{meff}
\end{equation}
and
\begin{equation}
{\tilde \mu}= \mu-2G_V \rho
\,\,,
\label{mueff}
\end{equation}
 Note also that, in principle, one should have two coupled gap equations 
for the two distinct flavors: $M_u = m_{u} - 2G_S(\langle {\bar u} u \rangle + \langle {\bar d} d \rangle)$ and  
$M_d = m_{d} - 2G_S(\langle {\bar d} d \rangle + \langle {\bar u} u \rangle)$ where $\langle {\bar u} u \rangle$ and  $\langle {\bar d} d \rangle$ 
represent the quark condensates which differ, due to the different electric charges. However, in the two flavor case, 
the different condensates contribute to $M_u$ and $M_d$ in a symmetric way and since $m_u=m_d=m$ one has $M_u=M_d=M$. 
The quantities $\phi=\phi_{vac}+\phi_{mag}+\phi_{med}$ and $\rho$ appearing in Eqs. (\ref {meff}) and (\ref{mueff}) are given by

\begin{equation}
\phi_{vac}=-
{\frac{MN_cN_f}{2\pi^2}}\left \{ \Lambda \epsilon_{\Lambda}
-\displaystyle {\frac{M^2}{2}}\ln{\left [ {\frac{(\Lambda+\epsilon_{\Lambda})^2}{M^2}} \right ]}
  \right \}\;\;,
\end{equation}
\begin{equation}
\phi_{mag}=-\displaystyle {\frac{MN_c}{2\pi^2}}\sum_{f=u}^d |q_f|B
\left [ \ln{\Gamma(x_f)}-{\frac{1}{2}}\ln{(2\pi)}+x_f-{\frac{1}{2}}(2x_f-1)\ln{(x_f)}      \right ]\;\;,
\end{equation}
\begin{equation}
\phi_{med}=\displaystyle {\frac{MN_c}{2\pi}}\sum_{f=u}^d \sum_{k=0}^{\infty}   \alpha_k(|q_f|B)\int_{-\infty}^{\infty} {\frac{dp_z}{2\pi}}{\frac{1}{E_{p}}}\left [ n_{p}(\tilde{\mu},T)+\bar n_{p}(\tilde{\mu},T) \right ] \;\;,
\end{equation}
and 
\begin{equation}
\rho = \displaystyle{\frac{N_c}{2\pi}} \sum_{f=u}^d \sum_{k=0}^{\infty}   \alpha_k(|q_f|B)\int_{-\infty}^{\infty} {\frac{dp_z}{2\pi}}
 \left[ n_{p}(\tilde{\mu},T)-\bar n_{p}(\tilde{\mu},T) \right] \,,
\end{equation}
where
\begin{equation}
\begin{array}{ccc}
n_p (\tilde{\mu},T)= \displaystyle{\frac{1}{1 + e^{(E_p - \tilde{\mu})/T}}} &{\rm and} & \bar{n}_p (\tilde{\mu},T)= \displaystyle{\frac{1}{1 + e^{(E_p+ \tilde{\mu})/T}}}\,,
\end{array} 
\end{equation}

\noindent represent, respectively, the Fermi occupation number for quarks and antiquarks.
 
At $T=0$ the relevant in-medium terms appearing in $\Omega$, $M$ and ${\tilde \mu}$ can be written as
\begin{equation}
{\Omega}_{med} = -\displaystyle{\frac{N_c}{4\pi^2}} \sum_{f=u}^d \sum_{k=0}^{k_{f,max}}   \alpha_k|q_f|B \Biggl \{ \tilde{\mu}k_F(k,B)-s_f(k,B)^2\ln{\left[ {\displaystyle\frac{\tilde{\mu}+k_F(k,B)}{s_f(k,B)}} \right]}\Biggr \}\;\;,
\label{Omega tau}
\end{equation}

\begin{equation}
\phi_{med}= \displaystyle\sum_{f=u}^d \sum_{k=0}^{k_{f,max}}\alpha_k {\frac{MN_c(|q_f|B)}{2\pi^2}}\ln{\left [ {\frac{\tilde{\mu}+k_F(k,B)}{s_f(k,B)}}    \right ]}\,\,,
\end{equation}
and 
\begin{equation}
 \rho = \sum_{f=u}^d \sum_{k=0}^{k_{f,max}} \alpha_k \frac{|q_f| B N_c}{2\pi^2}k_F(k,B)  \,\,,
\label{eq_rho_t0}
\end{equation}
\noindent
where $k_F(k,B)$ represents the Fermi momentum,  $k_F=\sqrt{{\tilde \mu}^2-s_f(k,B)^2}$ , and $s_f(k,B) = \sqrt{M^2+2|q_f|kB}$.  The upper Landau
level (or the nearest integer) is defined by
\begin{equation}
k_{f,max} =\displaystyle \left \lfloor {\frac{\tilde\mu^2-M^2}{2|q_f|B}} \right \rfloor .
\label{kmax}
\end{equation}
To obtain numerical results we must now fix the model parameters and, as usual, the cut-off value together the other parameters $G_S$, $m$ are chosen to 
reproduce the phenomenological values \cite{buballa} for the pion mass ($m_\pi \simeq 140 \, {\rm MeV}$), the pion decay constant ($f_\pi \simeq 93\, {\rm MeV}$) and the quark condensate 
($\langle {\bar \psi} \psi \rangle^{1/3} \simeq 250 \, {\rm MeV}$).  Here, we consider the set $\Lambda = 590 \,  {\rm MeV}$, $G_S \Lambda^2 = 2.435$ and $m = 6.0\,  {\rm MeV}$. 

Fixing $G_V$ poses and additional problem since this quantity should be fixed using the $\rho$ meson mass which, in general, happens to be higher than the maximum energy scale set by $\Lambda$. At present,  the vector term  coupling  $G_V$ cannot be determined from experiments and lattice QCD simulations (LQCD) but  eventually, the combination of neutron star observations and the energy scan  of the  phase-transition signals 
at FAIR/NICA may provide us some hints on its precise numerical value.   Therefore, at the present stage the vector coupling $G_V$ is usually taken to be  a free parameter whose  
accepted values  lie within the range $0.25\,G_S-0.5\,G_S$ \cite{gv1,gv2,sugano}. It is worth to point out that the explicit use of $G_V$ within the NJL model can be avoided provided that  the evaluations be performed beyond the large-$N_c$ (or MFA) level. In this case two loop (exchange like) terms bring finite $N_c$ corrections to the pressure such as $(G_S/N_c) \rho^2$ so that the same type of physics can be observed with one less parameter \cite {tulio}. 

\section{Cascade of First Order Chiral Transitions with Two Coexisting Densities}

Within the usual first order transition scenario    the free energy develops a pair of degenerate global minima    defining two distinct values for the effective quark masses as well as for the quark number density. Usually, when $B=0$,  only one first order chiral transition takes place  so that, for a given temperature, a unique value for the coexistence chemical potential exists. However, as already emphasized, the presence of a magnetic field  causes an oscillatory behavior which  may induce more exotic patterns 
like the appearance of a cascade of  first order transitions such as the ones studied 
 in Refs. \cite {ebert, Allen:2013lda, Grunfeld:2014qfa} where only the case of vanishing temperatures, in the absence of repulsion, has been considered. One of the main outcomes of these applications is that, for some phenomenologically acceptable parametrizations, several transitions are needed to move from the vacuum phase  to the (approximately) chirally symmetric  phase. More recently, Allen, Pagura and Scoccola \cite {pablo2} have observed a similar situation when considering a a generalized version of the two flavor NJL model, which includes a $G_V$ term, but again at vanishing temperatures. In this section we will extend the investigation performed at $G_V \ne 0$ to the finite temperature case  to gain extra insight on the effects of $B$ and $G_V$ by exploring the phase diagram in  planes such as $T-\mu$, $T-\rho_B$, and $P-T$. This exercise  will allow us not only to review some of the main aspects related to the appearance of a cascade of first order phase transitions as the ones studied in Refs. \cite {ebert, Allen:2013lda, Grunfeld:2014qfa, pablo2} but also to explore more realistic finite temperature situations.

\begin{figure}[htp]
\centering
\includegraphics[width=0.4\textwidth]{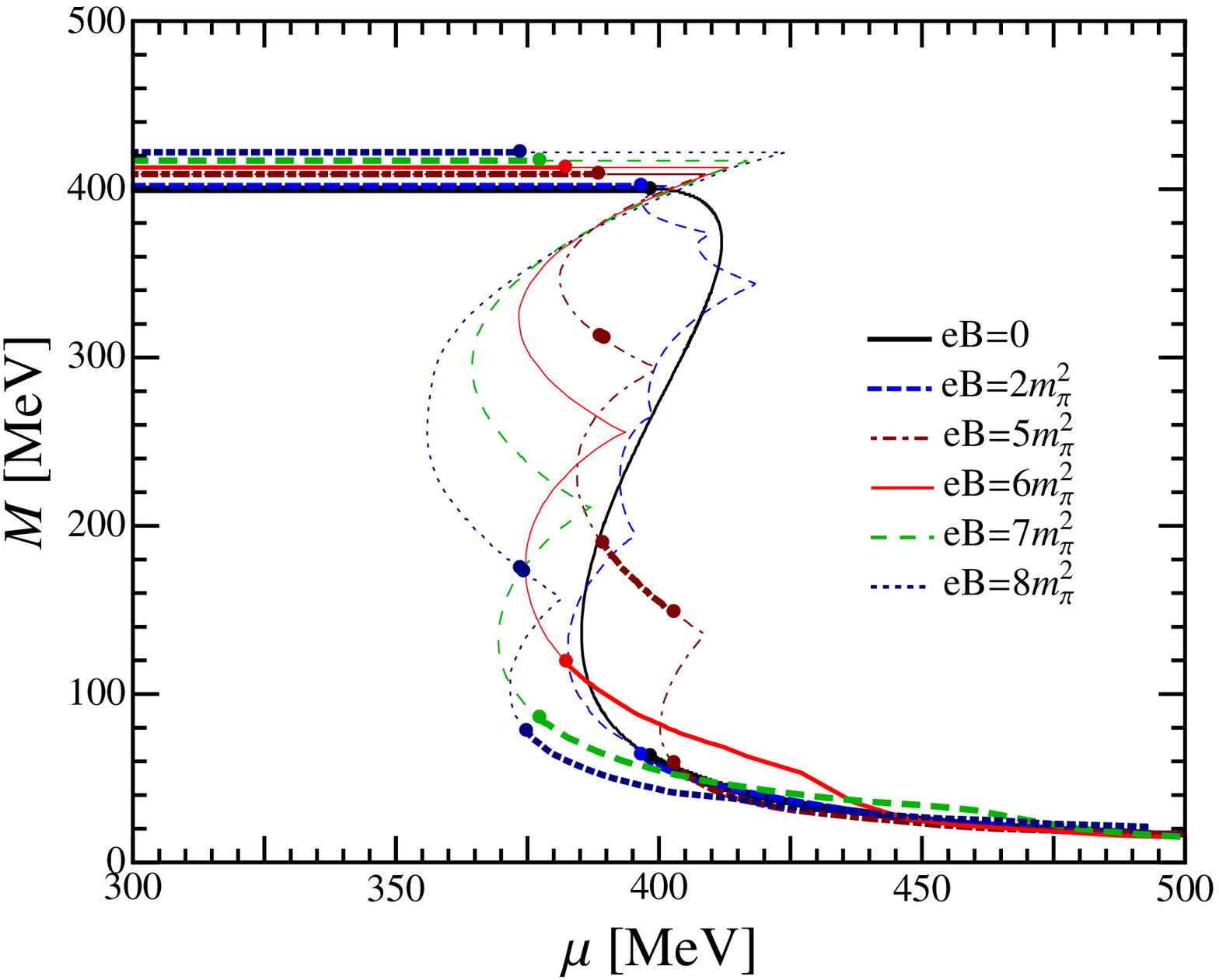} 
\hspace{10pt}
\includegraphics[width=0.4\textwidth]{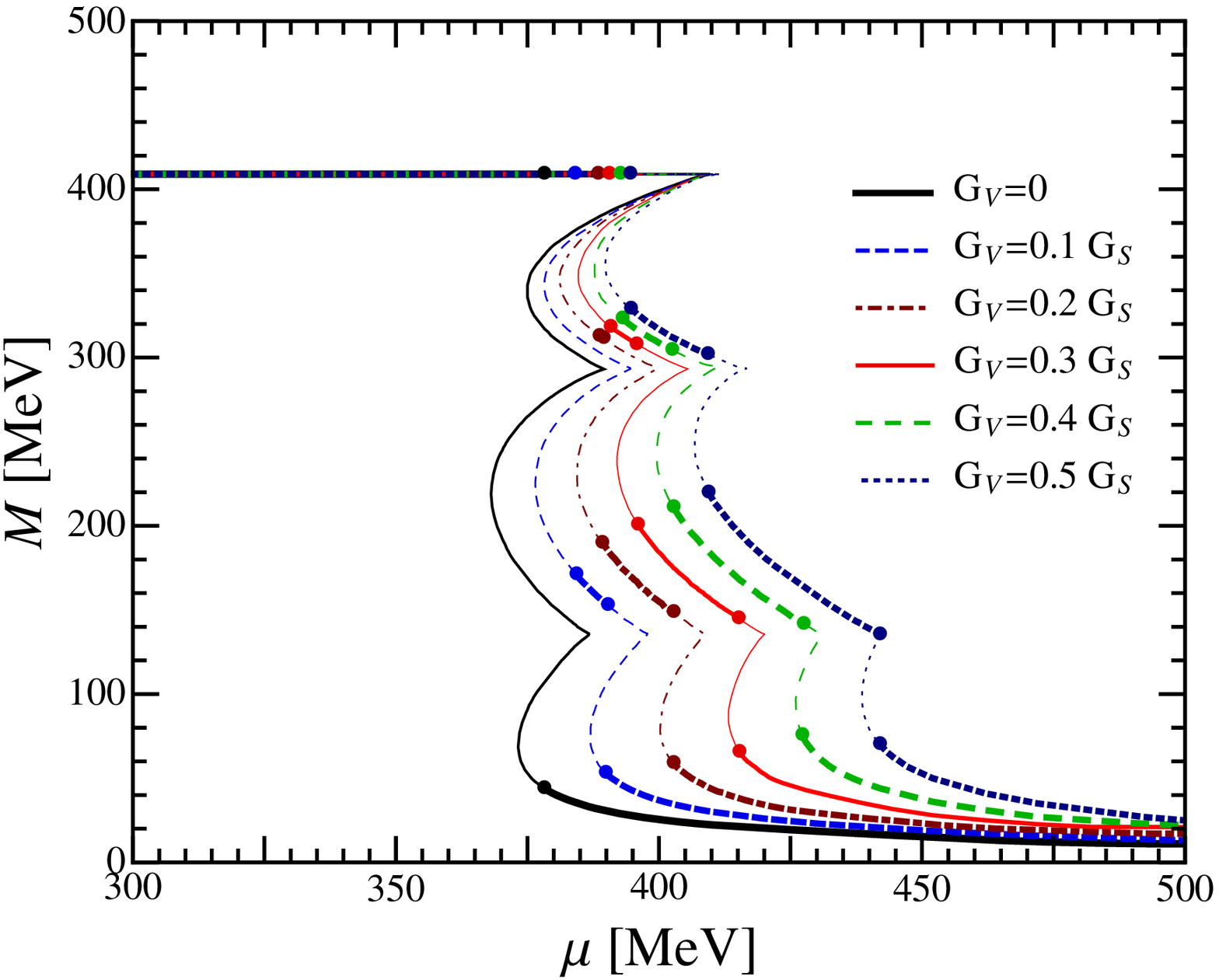} 
\caption{ Cascade of first order  transitions. Left panel: The effective quark mass, $M$, at $T=0$, as a function of $\mu$ for 
different values of $eB$ at $G_V=0.2\,G_S$. 
Right panel: The effective quark mass, $M$, at $T=0$, as a function of $\mu$ for different values of $G_V$ at $eB=5\, m_\pi^2$. 
In both cases the thick lines represent stable solutions to the gap equation.}
\label{fig1}
\end{figure}

\par
Let us first recall that, within this model, the order parameter associated with the chiral transition is  $ (\langle {\bar u} u \rangle + \langle {\bar d} d \rangle)/2$ which, in our case, is directly related to the effective mass as Eq. (\ref {meff}) indicates. Therefore, it is instructive to analyze the effects of $B$ and $G_V$ over $M$ since this quantity also determines the behavior of the associated EoS. The left panel of Fig. \ref {fig1}, which was originally obtained in Ref. \cite {Denke:2013gha},  displays the effective quark mass, at $T=0$, as a function of $\mu$ for $G_V=0.2\, G_S$ and different values of the magnetic field.  The figure indicates that, in the vacuum, the value of $M$ tends to increase with $B$ which is in accordance with the magnetic catalysis effect \cite {MC}. Also, due to the filling of Landau levels, one observes the typical de Haas-van Alphen oscillations which are more pronounced for small values of $B$. 
Only the segments of the curves where $dM/d\mu < 0$ correspond to energetically favored gap equation solutions 
and in the present case ($G_V=0.2 \, G_S$) we observe that for $eB=5\, m_\pi^2$ and $eB=8\, m_\pi^2$ some of these solutions are stable leading to intermediate transitions.  For example, at $eB =5 \, m_\pi^2$ one sees a first transition at $\mu=388.55 \, {\rm MeV}$ when the mass jumps from $409\, {\rm MeV}$ to $ 313 \, {\rm MeV}$.  This is followed by another transition from $M=312.6 \, {\rm MeV}$ to  $M=190 \, {\rm MeV}$ at $\mu=390 \, {\rm MeV}$. To complete the  ``cascade" of (three)  first order phase transitions one observes a final transition from $M=150 \, {\rm MeV}$  to $59\,{\rm MeV}$ at $\mu=402.65 \, {\rm MeV}$. 
The numerical results illustrated by the figure  clearly display the effective  mass oscillatory behavior showing that at relatively weak magnetic fields 
we observe many oscillations of quark mass values (due to the many Landau levels available).  When the field becomes stronger, the quantity of 
oscillations in the quark effective mass is reduced since there are less Landau levels available. The left panel also shows that, for a fixed value of $G_V$, the transition to the chiral phase (lowest mass value) occurs at lower chemical potential coexistence values as $B$ increases in accordance with the IMC phenomenon. The right panel shows that, for a fixed $B$, the transition to the chiral phase  occurs at higher chemical potential coexistence values as $G_V$ increases (for a complete discussion on these issues the reader is referred to Ref. \cite  {Denke:2013gha}).
For our present purposes it is important to note how the insertion of a repulsive vector coupling $G_V$ between quarks  brings stability ($dM/d\mu \le 0$) to the 
system and occasionally promotes these magnetic intermediary density states to stable ones as the right panel shows.
\begin{figure}[htp]
\centering
\includegraphics[width=0.4\textwidth]{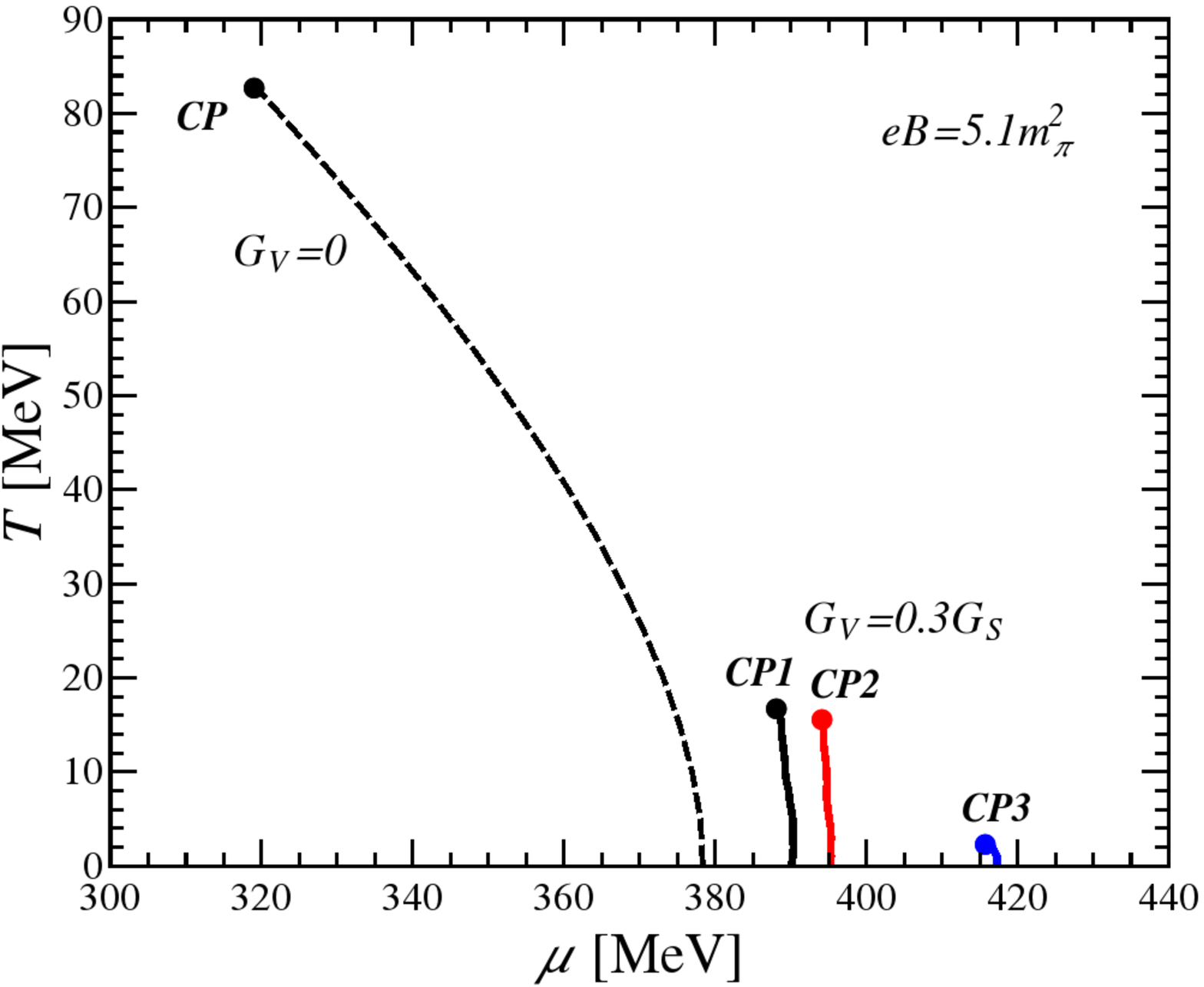}
\hspace{10pt} 
\includegraphics[width=0.395\textwidth]{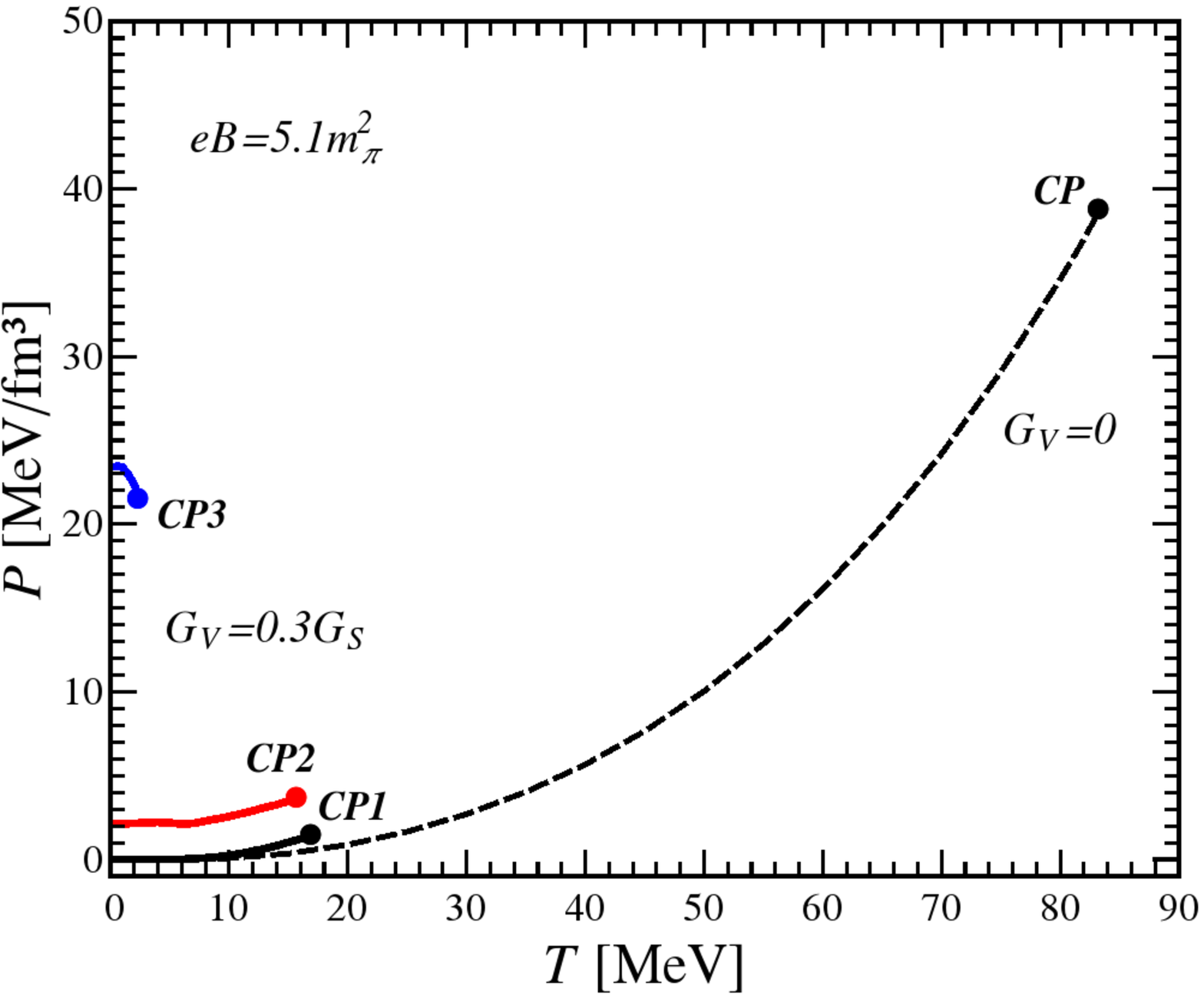} 
\caption{Quark matter phase diagrams in the $T-\mu$ plane (left panel) and in the $P-T$ plane (right panel) 
for  $eB=5.1\,m_{\pi}^2$ and  $G_V=0.3\,G_S$. In both figures the continuous thin line represents the case with $G_V=0$ 
for comparison. When $G_V=0.3\,G_S$ a cascade of three first order transitions appear at very low temperatures. Each first order transition line terminates at the indicated critical point. }
\label{fig2}
\end{figure}

\par
Figure \ref{fig2}  shows the $T-\mu$ and $P-T$ phase diagrams for the case of a field strength\footnote{The choice of this particular value will become clear in the sequel.} $eB=5.1 \,m_{\pi}^2$ and a vector coupling magnitude of $G_V=0.3\,G_S$, which is the value recently suggested by Sugano et al. \cite {sugano}, 
 compared to the 
$G_V=0$ case. At  temperatures close to zero, one observes a splitting of the $G_V=0$ first order transition line into three lines occurring at different $\mu$ values as one could expect from the discussion related to  figure \ref {fig1}. As already emphasized this exotic scenario may also appear in the $G_V=0$ case provided that one uses a different parametrization for $G_S$  as discussed in Refs. \cite {ebert, Allen:2013lda, Grunfeld:2014qfa}. Here, we have instead adopted a rather canonical  value for $G_S$ so that only one transition occurs if $G_V=0$. This standard  choice is well suited to achieve our goal since the role played by the vector channel itself can be further highlighted. 
The left panel of this  figure shows that, as expected \cite {fukushima}, $G_V$ 
weakens the first order transition and shifts the coexistence chemical potential 
to high values. Note also that for $G_V \ne 0$   the third transition 
(thick continuous line) is quickly washed out for temperatures higher 
than $\approx 4.75 \, {\rm MeV}$. The right panel shows the phase diagram in the 
physically more intuitive $P-T$ plane. It is interesting to note that the transition 
terminating at CP3, which is associated with the Landau level jump $k_d = 1 \to 2$, 
has a negative $\Delta P/\Delta T$ slope (related the Clausius-Clayperon equation)  
while the ones terminating at $CP1$ and $CP2$, associated with $k_d = 0 \to 0$, have a positive slope just like the one usually observed when $G_V=0$. Therefore, apart from the usual ``liquid-gas" type of transition (positive slope) it appears that the combined  presence of $G_V$ and $B$ may also induce a transition of the ``solid-liquid" type observed in the water phase diagram (negative slope) which, as far as we know, has not been observed before within QCD motivated models.

\begin{figure}[htp]
\centering
\includegraphics[width=0.4\textwidth]{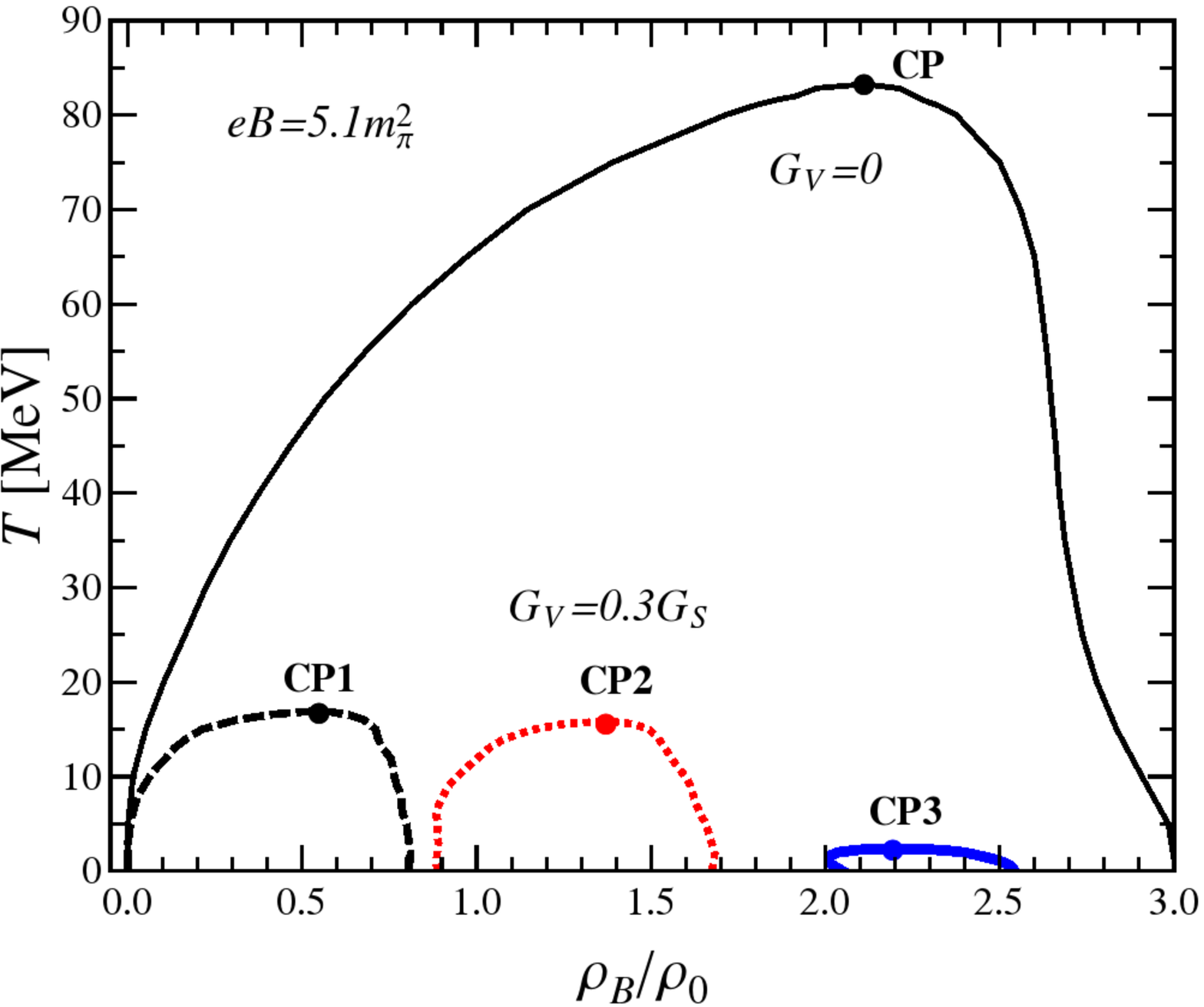} 
\hspace{10pt}
\includegraphics[width=0.4\textwidth]{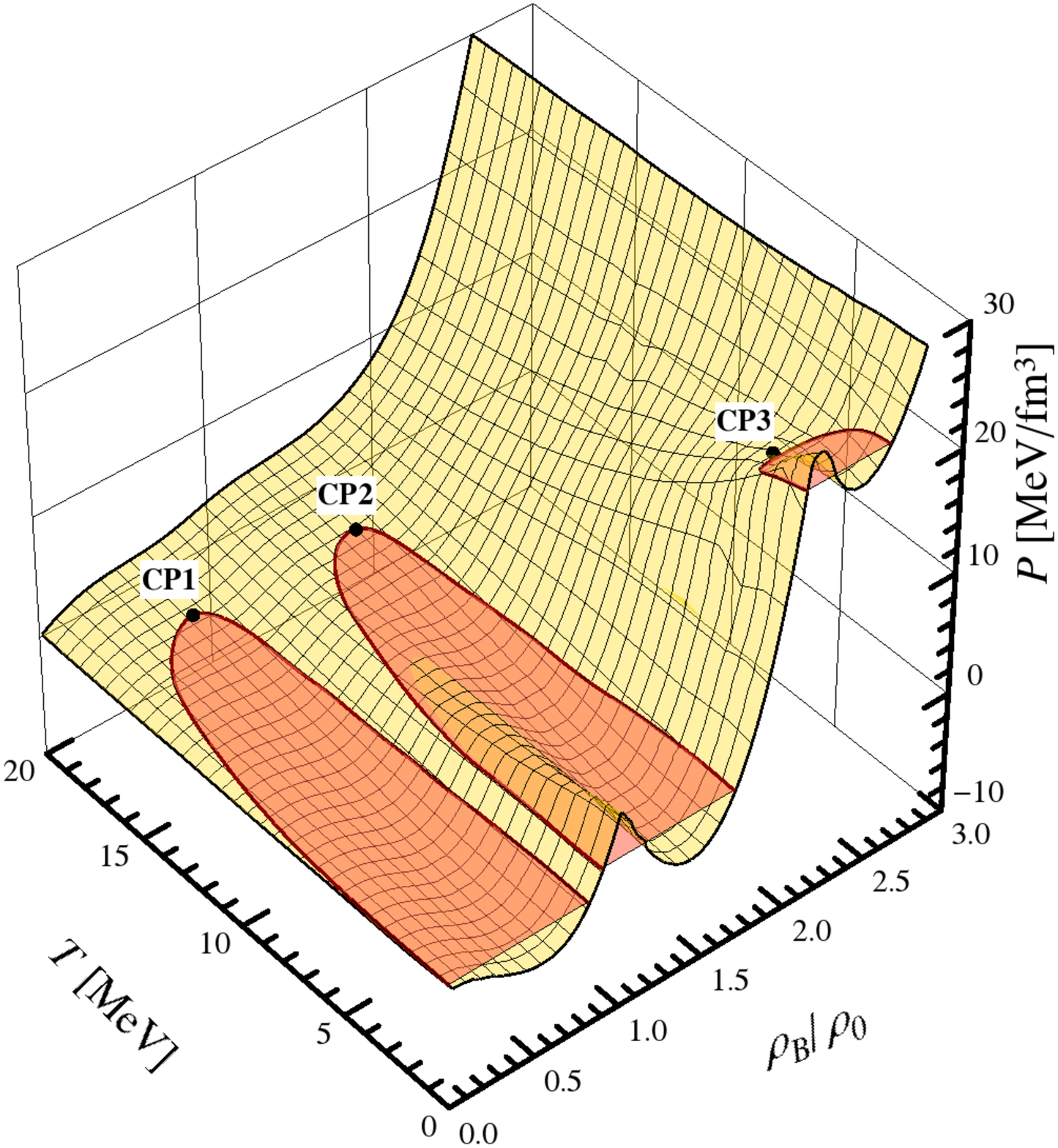}
\caption{Left Panel: Coexistence phase diagram in the $T-\rho_B$ plane for $eB=5.1 \,m_{\pi}^2$ and  $G_V=0.3\,G_S$. The case $G_V=0$ (thin continuous line) is shown for comparison. Right panel: The  phase diagram in the $P-T-\rho_B$ space. In both panels it is  
possible to distinguish the three independent first order phase transitions defined by the coexistence of two distinct densities  at the same pressure.}
\label{fig3}
\end{figure}

Next, in the left panel of Fig. \ref {fig3} we  show the  phase diagram in the $T-\rho_B$ plane which could not be  analyzed in the previous applications at $T=0$ \cite {ebert, Allen:2013lda, Grunfeld:2014qfa, pablo2}. The figure shows that for the chosen $G_V$ and $B$ a values a shrinkage of the coexistence regions also takes place. At $T=0$ the figure shows a coexistence between the following pairs of densities: $\rho_B = 0$ and $\rho_B=0.8\, \rho_0$ (at $\mu=388.55 \, {\rm MeV}$); $\rho_B = 0.9 \, \rho_0$ and $\rho_B=1.7\, \rho_0$ (at $\mu=390 \, {\rm MeV}$); and $\rho_B = 2 \, \rho_0$ and $\rho_B=2.55\, \rho_0$ (at $\mu=402.65 \, {\rm MeV}$) whereas for the $G_V=0$ case the dilute phase occurs at vanishing density and the dense phase occurs at $\rho_B = 3 \, \rho_0$. As usual the baryonic density is defined as $\rho_B=\rho/3$ while $\rho_0= 0.17\, {\rm fm}^{-3} $. 
According to Refs \cite {jorgen,andre} one can then expect that the surface tension between the coexistence phases will be lower when $G_V \ne 0$. The right panel of Fig. \ref {fig3} shows a three dimensional plot displaying the $P-T-\rho_B$ phase diagram with the Andrews isotherms that define the equation of state. This figure  indicates  that the  cascade of three first order transition observed so far    always refer to the coexistence of  two phases occurring at distinct pressures.

Up to this point we have seen that, due to the
LL filling procedure,  chiral symmetry restoration may take place
via successive first order transitions between different magnetized phases. From the free energy perspective this succession happens because, depending on the values of the couplings, the presence of a magnetic field may induce the appearance of  more than two minima so that by varying a control parameter, such as $\mu$, one may observe multiple transitions. This scenario can happen either when $G_V=0$ and $G_S$ is relatively weak (leading to effective quark masses such as $M \approx 200 \, {\rm MeV}$)    \cite {ebert, Allen:2013lda, Grunfeld:2014qfa} or when $G_V \ne 0$  but $G_S$ leads to more standard $M$ values as we have shown.  At this point it is crucial to note that, with the values of $G_V$ and $B$ considered so far, we have only observed the occurrence of two  degenerate global minima signaling  the usual type of  first order phase  phase transition for a given value of $\mu$ while  any eventual extra minima remain local. Then, at another chemical potential value a  global minimum turns into a  local one and vice versa leading us to  observe a cascade of transitions where only 
two densities coexist for a given value of $\mu$. The results found in this section  suggest that perhaps there are certain values of $G_V$ at  which  three degenerate  global minima will emerge signaling the coexistence of {\it three} different phases at the same $T,\mu$ and $B$ values  as  we shall discuss next.

\section{First Order Chiral Transitions with Three Coexisting Densities}

\begin{figure}[htp]
\centering
\includegraphics[width=0.4\textwidth]{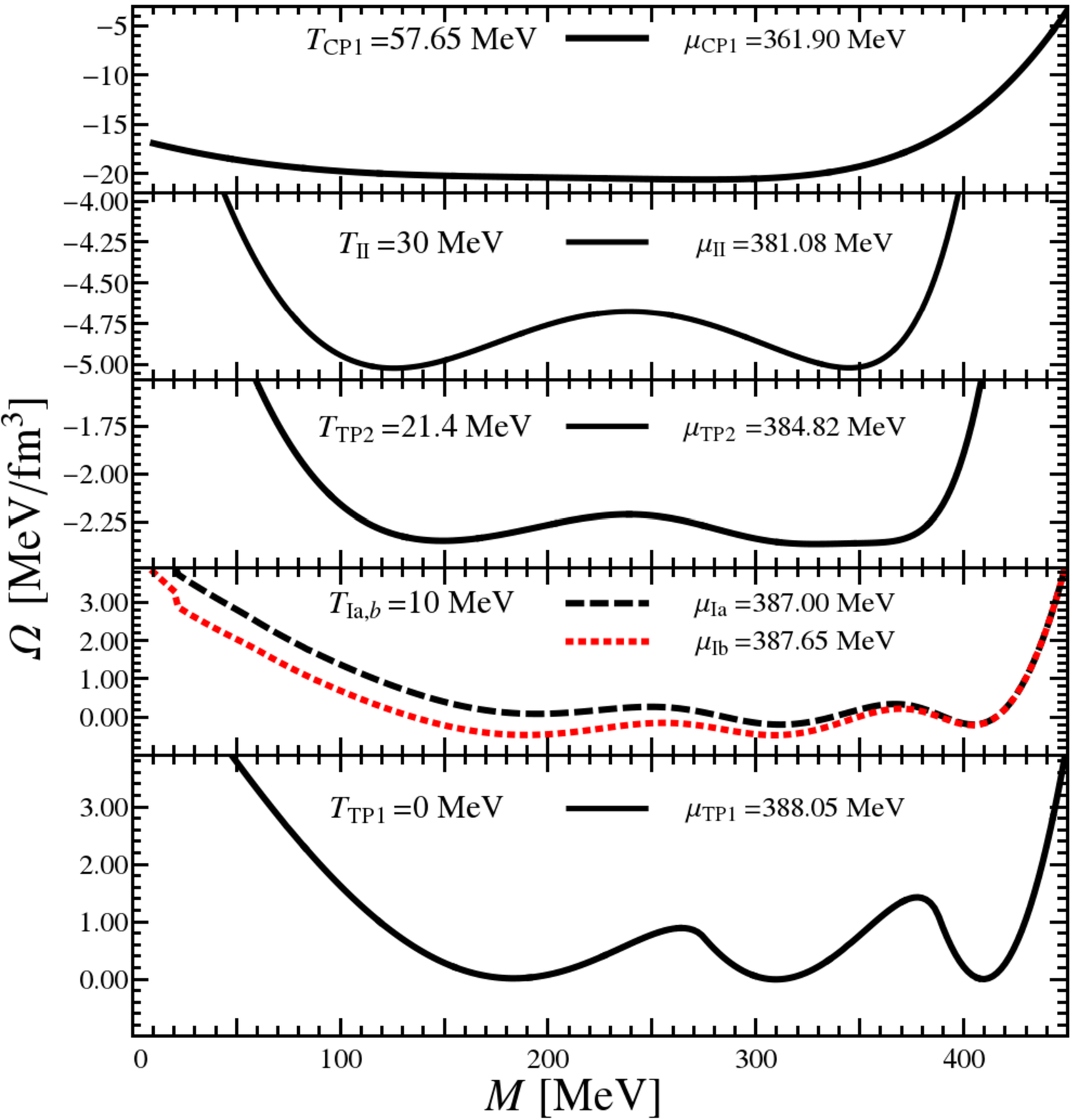} 
\hspace {10pt}
\includegraphics[width=0.4\textwidth]{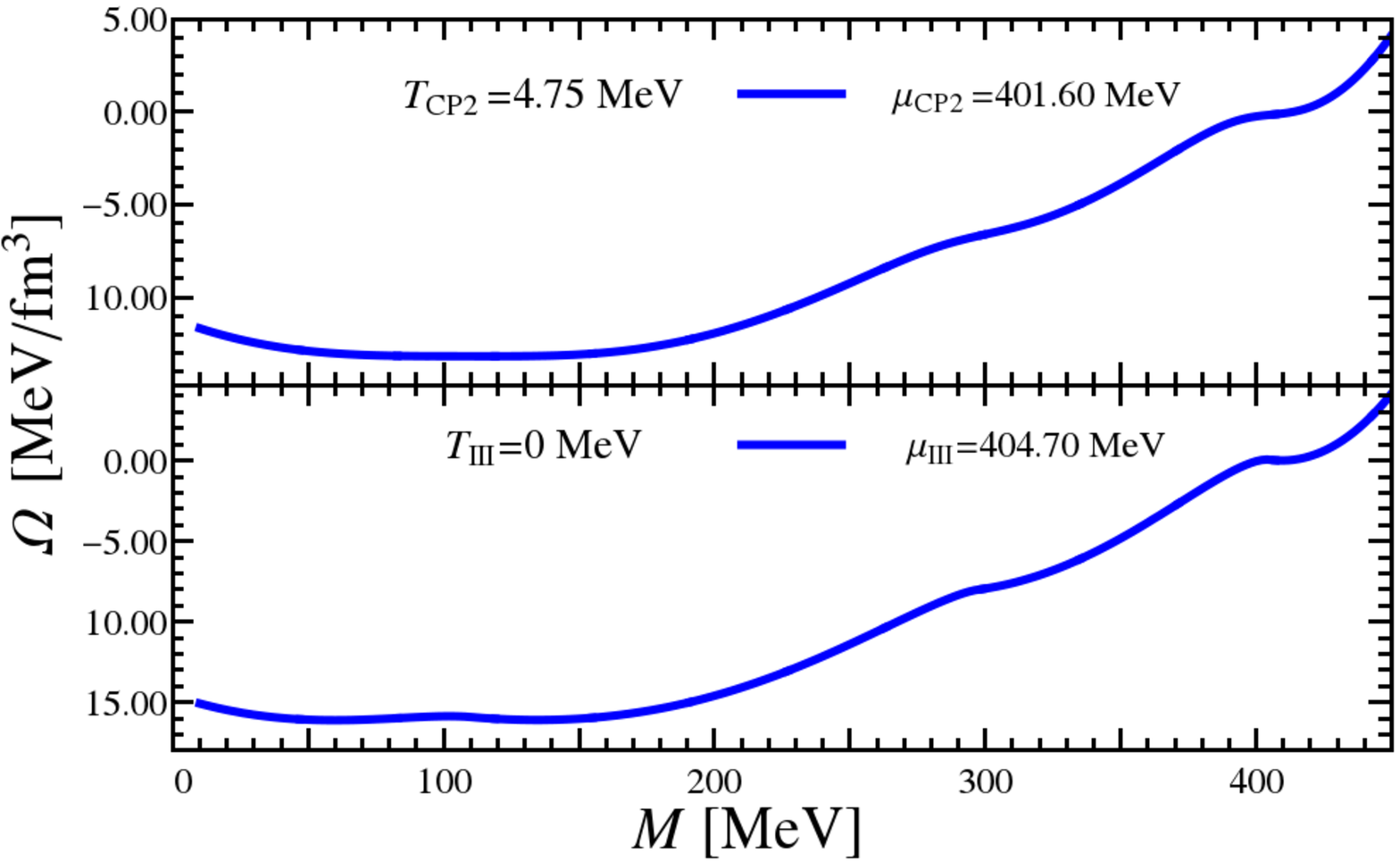} 
\caption{Thermodynamic potential at $G_V=0.2\, G_S$ and $eB=5.1\, m_\pi^2$ for some selected temperatures. Left panel: at $T=0$ the potential is triply degenerate at the coexistence chemical potential value $\mu_{TP1}=388.05\, {\rm MeV}$. At $T=10 \, {\rm MeV}$ the potential is doubly degenerate at two different coexistence chemical potential values, $\mu_{Ia}=387\, {\rm MeV}$ and $\mu_{Ib}=387.65 \, {\rm MeV}$. At $T=21.4\, {\rm MeV}$ it becomes triply degenerate again, at $\mu_{TP2}=384.82\, {\rm MeV}$  while at $T=30 \, {\rm MeV}$ it is doubly degenerate at $\mu_{II}=381.08\, {\rm MeV}$. Right panel: 
}
\label{fig4}
\end{figure}

The previous discussion shows that the $B$ and $G_V$ values considered so far  produce a cascade of first order transitions where, at a given coexistence $\mu$ value, only a pair of  degenerate (global) minima  coexist with other (local) minima. However, for other parameter values, it may be energetically preferable  that one of these local minima  becomes a global one so that the ground state is triply degenerated.  With this motivation let us now  scan $G_V$ and $eB$ around the values $G_V=0.2 \, G_S$ and $eB=5 \, m_\pi^2$. 
Fig. \ref {fig4} shows the thermodynamic potential evaluated at $eB=5.1\,m_{\pi}^2$ and $G_V=0.2\,G_S$ at various temperatures. To facilitate the understanding of the unusual phase diagrams which  appear in the sequel  let us discuss the results shown in this figure  with some detail.
Starting with the $T=0$ case one observes (left panel) three coincident minima occurring at the same null pressure coexistence chemical potential value $\mu_{TP1}=388.05\,{\rm MeV}$ where the subscript stands for ``triple point 1".

At $T=10\,{\rm MeV}$ this triple coexistence at the same chemical potential no longer survives and one observes
the dissociation into two separate first order transitions occurring at two distinct coexistence chemical potential 
values $\mu_{Ia}=387\,{\rm MeV}$ and $\mu_{Ib}=387.65\,{\rm MeV}$, respectively which signals a ``cascade" of two subsequent first order phase transitions. At a higher temperature, near $T_{TP2}=21.4\,{\rm MeV}$, three degenerate minima are again observed and one can ascribe another triple point taking place at $\mu_{TP2}=384.82 \,{\rm MeV}$. Above this temperature, we observe the usual behavior of the chiral with only one first order transition which
evolves to the critical point as $T$ increases. The right panel of Fig. \ref {fig4} shows a final transition line starting at $T=0$ and $\mu_{III}=404.70\,{\rm MeV}$ and terminating at the CP located at $T_{ CP2}=4.75\, {\rm MeV}$ and $\mu_{ CP2}=401.60\, {\rm MeV}$. 

\begin{figure}[htp]
\centering
\includegraphics[width=0.4\textwidth]{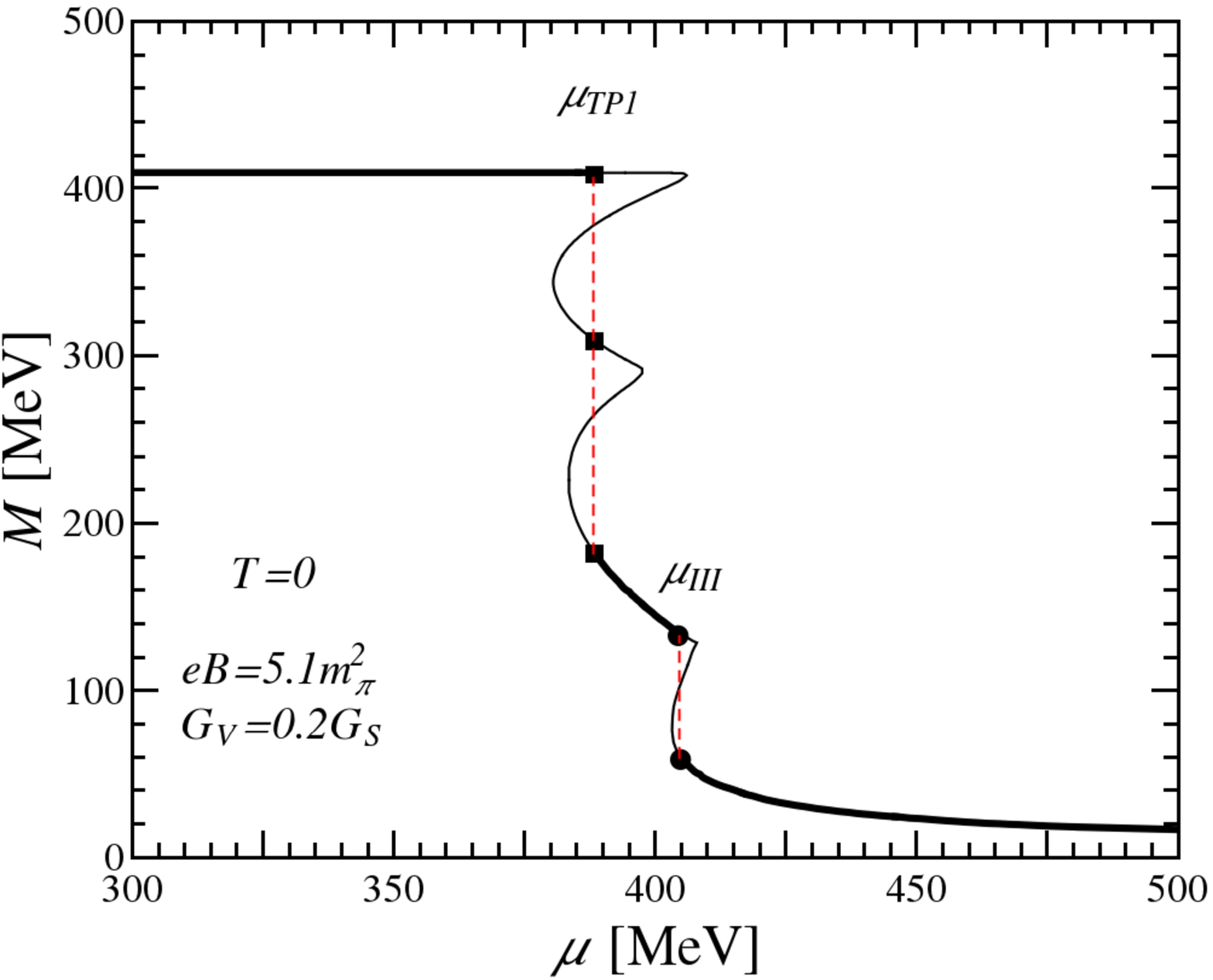} 
\hspace{10pt}
\includegraphics[width=0.4\textwidth]{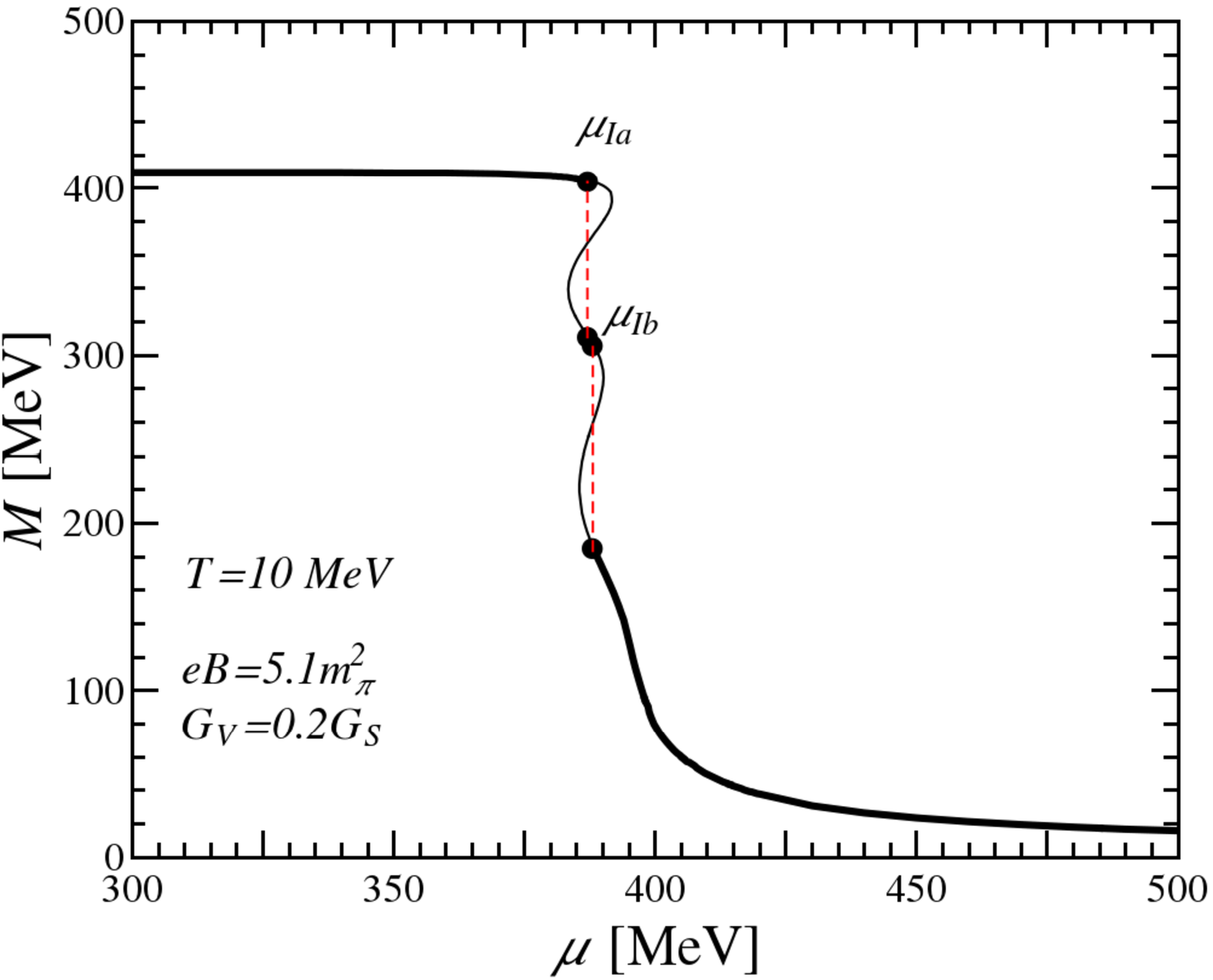} 
\includegraphics[width=0.4\textwidth]{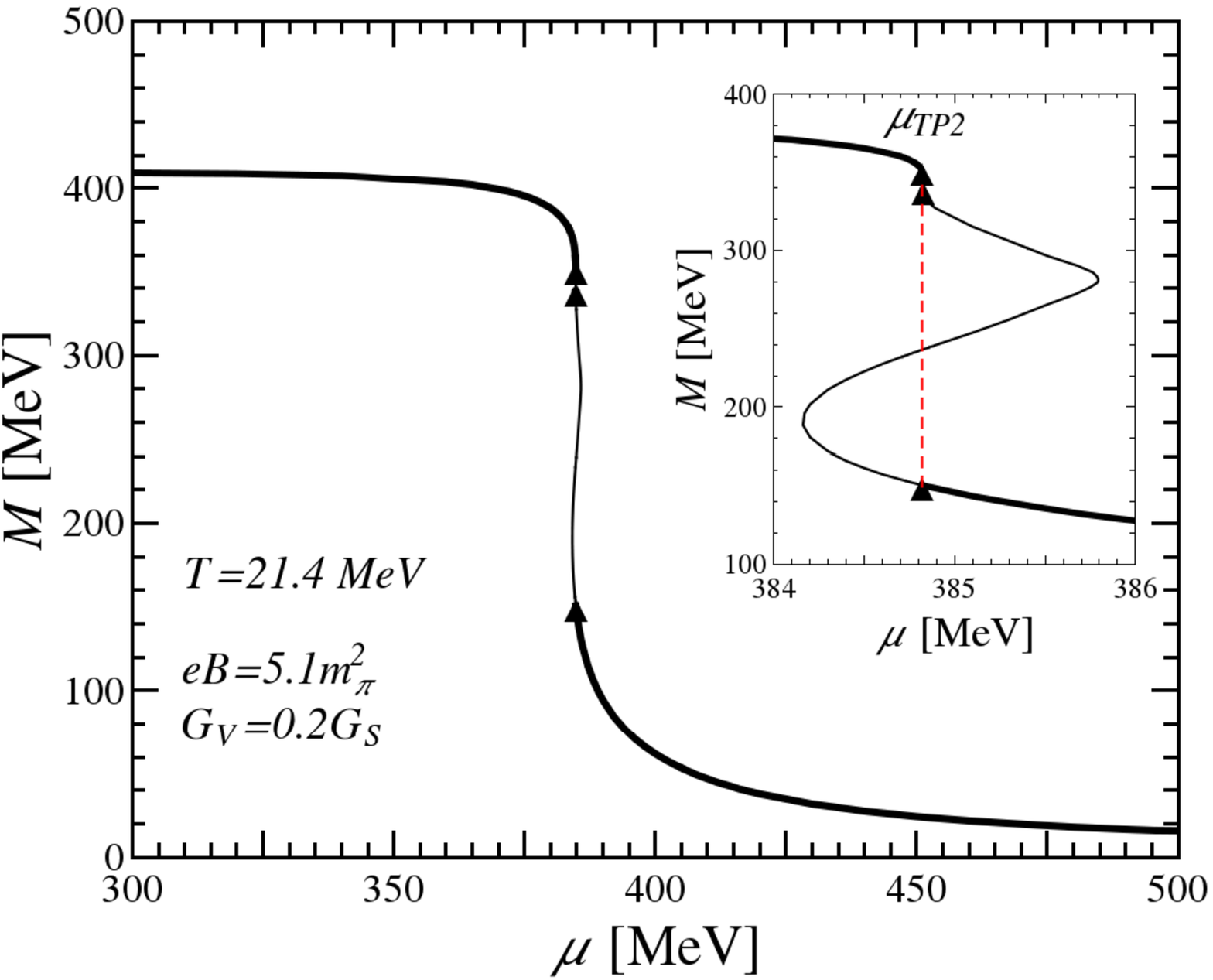} 
\hspace{10pt}
\includegraphics[width=0.4\textwidth]{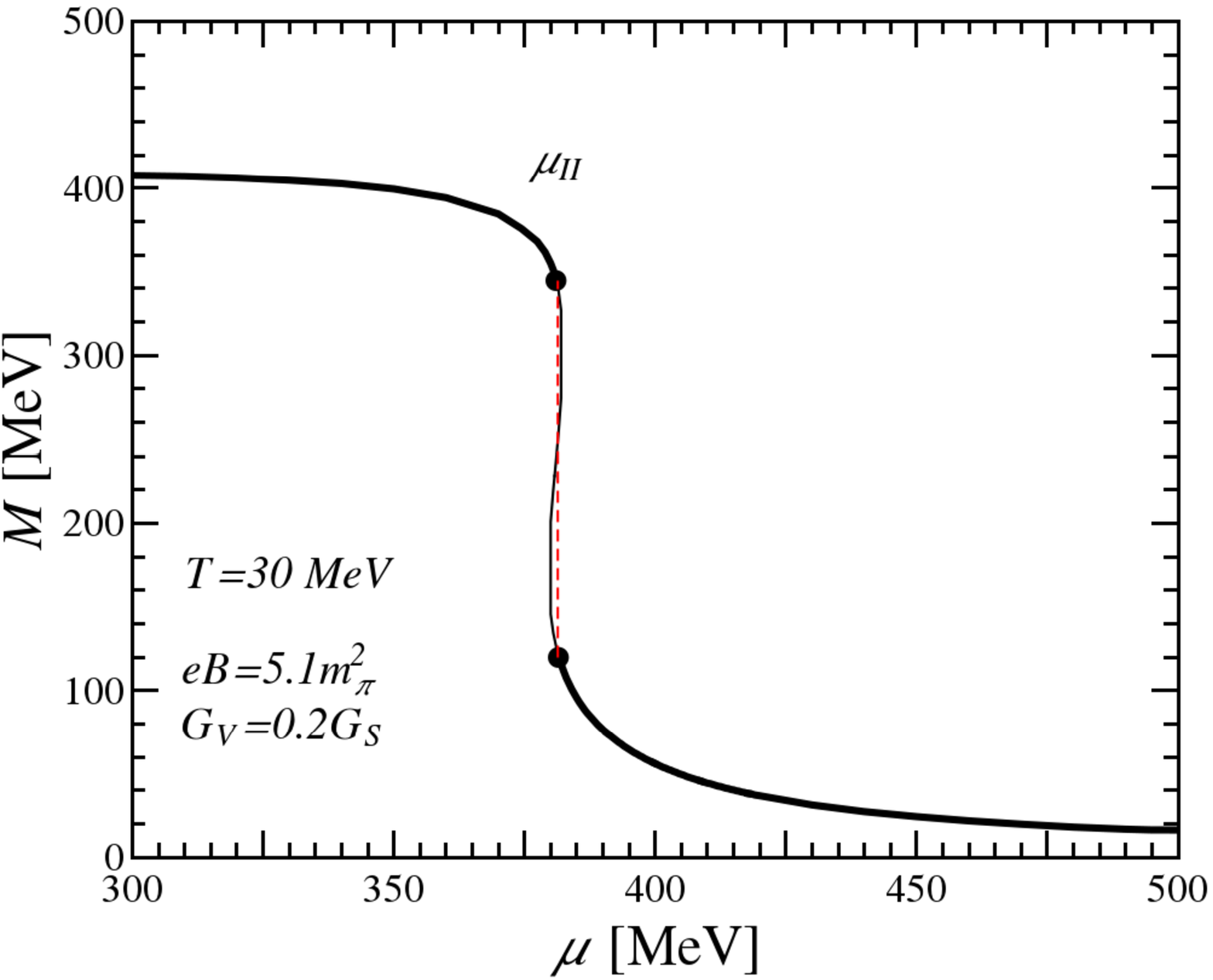} 
\caption{Sequence of plots showing first order chiral transitions for  $eB=5.1\,m_{\pi}^2$ and  $G_V=0.2\,G_S$. The squares and triangles  indicate triply degenerate transitions while the dots indicate doubly degenerate ones.  }
\label{fig5}
\end{figure}

Notice that in the previous section we have deliberately shown results for the cases $G_V=0.2 \, {\rm MeV}$ 
and $eB=5 \, {\rm MeV}$ as well as for  $G_V=0.3 \, {\rm MeV}$ and $eB=5.1 \, {\rm MeV}$ but not for the present choice, $G_V=0.2 \, {\rm MeV}$ and $eB=5.1 \, {\rm MeV}$, which induces the appearence of triple points. 
As already discussed the quark effective mass  is directly related to the order parameter so that the discussion can be further clarified by investigating how this quantity varies with $\mu$ for the $T,G_V$ and $B$ values considered in Fig. \ref {fig4}. With this aim we present Fig. \ref {fig5} where  one may  observe the transitions discussed in connection with the former figure from a different perspective. The squares locate the first triple point, the triangles locate the second triple point while the ordinary type of first order transitions is denoted by the dots.

From the sequence shown, it becomes evident how the triple point $TP1$, related to three different mass values (marked by the squares), 
determines the same chemical potential $\mu_{TP1}=388.05\,{\rm MeV}$ when $T=0$. Note that the the three mass values, satisfying 
the gap equation, differ by an approximately equal amount (close to $100\, {\rm MeV}$). Still at $T=0$ but at a higher 
chemical potential value one observes the occurrence of another (ordinary) transition signaled by two different mass 
values  (marked by the dots) at $\mu_{III}= 404.70 \, {\rm MeV}$.  Then,  rising the temperature  to $T=10\,{\rm MeV}$  one sees that the   triple coexistence, observed at $T=0$, decouples into two ordinary first order transitions 
occurring at two different, but yet very similar, chemical potentials given by $\mu_{Ia}=387\, {\rm MeV}$ and 
$\mu_{Ib}=387.65 \, {\rm MeV}$. At this temperature the first order line starting at $\mu_{III}$ and $T=0$ 
has already vanished. Note also that the low mass value occurring at $\mu_{Ia}$ and the high mass value 
occurring $\mu_{Ib}$ are  almost identical (the difference amounts to about $5\, {\rm MeV}$). 
Near $T=21.4\,{\rm MeV}$, these distinct first order lines start to approach each other and one observes another 
triple point ($TP2$) at $\mu_{TP2}$ marked by the triangles. In this case the high and the intermediate mass 
values become almost identical but  differ substantially from  the low value (by almost $200 \, {\rm MeV}$). 
Above this temperature, as the $T=30 \, {\rm MeV}$ panel suggests,  only the ordinary scenario takes 
place until the   first order transition line ends at  a critical point, $CP1$, located  
at $T_{CP1}= 57.65\, {\rm MeV}$ and $\mu_{CP1}=361.90\, {\rm MeV}$.
\par
\begin{figure}[htp]
\centering
\includegraphics[width=0.397\textwidth]{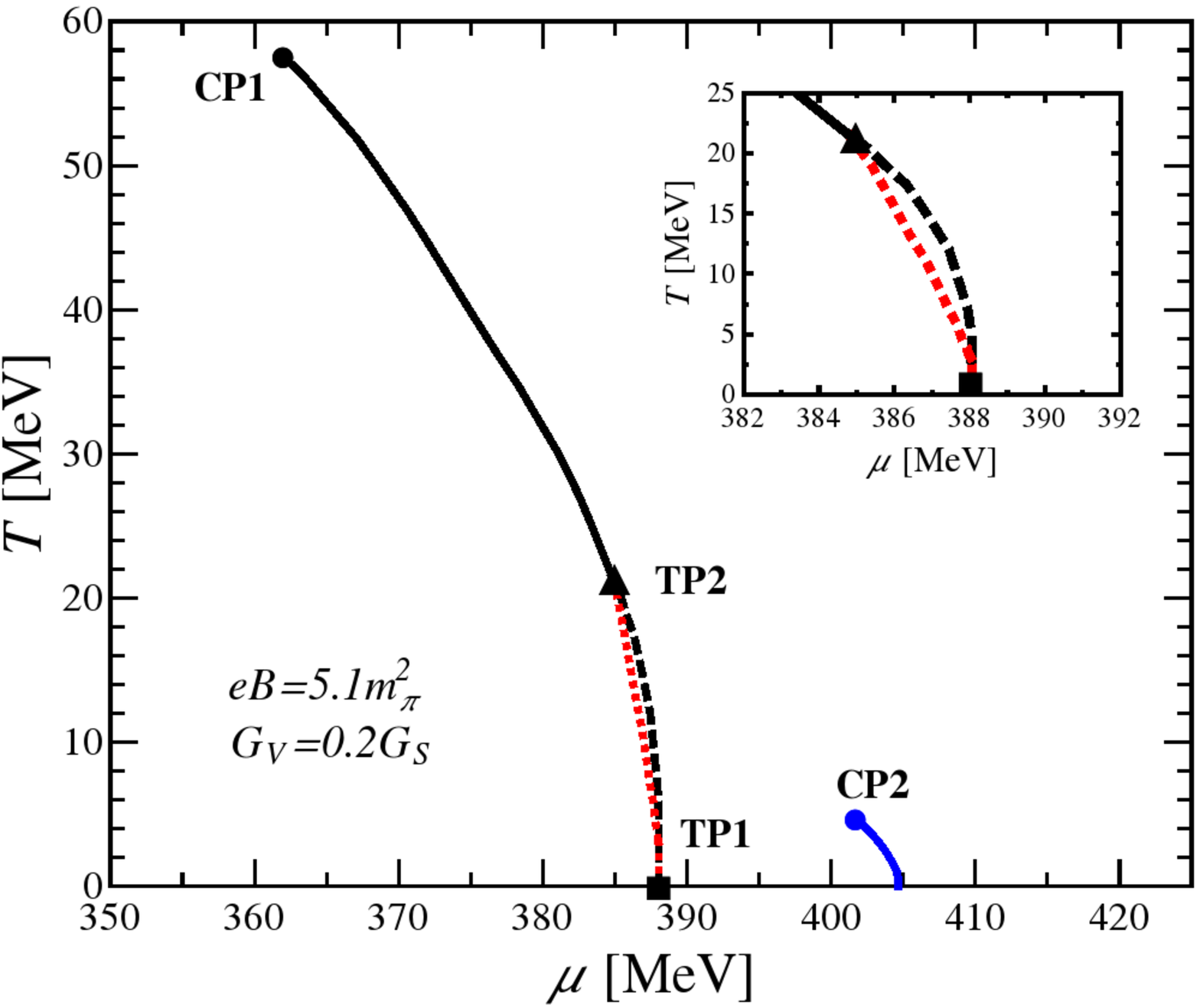} 
\hspace{10pt}
\includegraphics[width=0.4\textwidth]{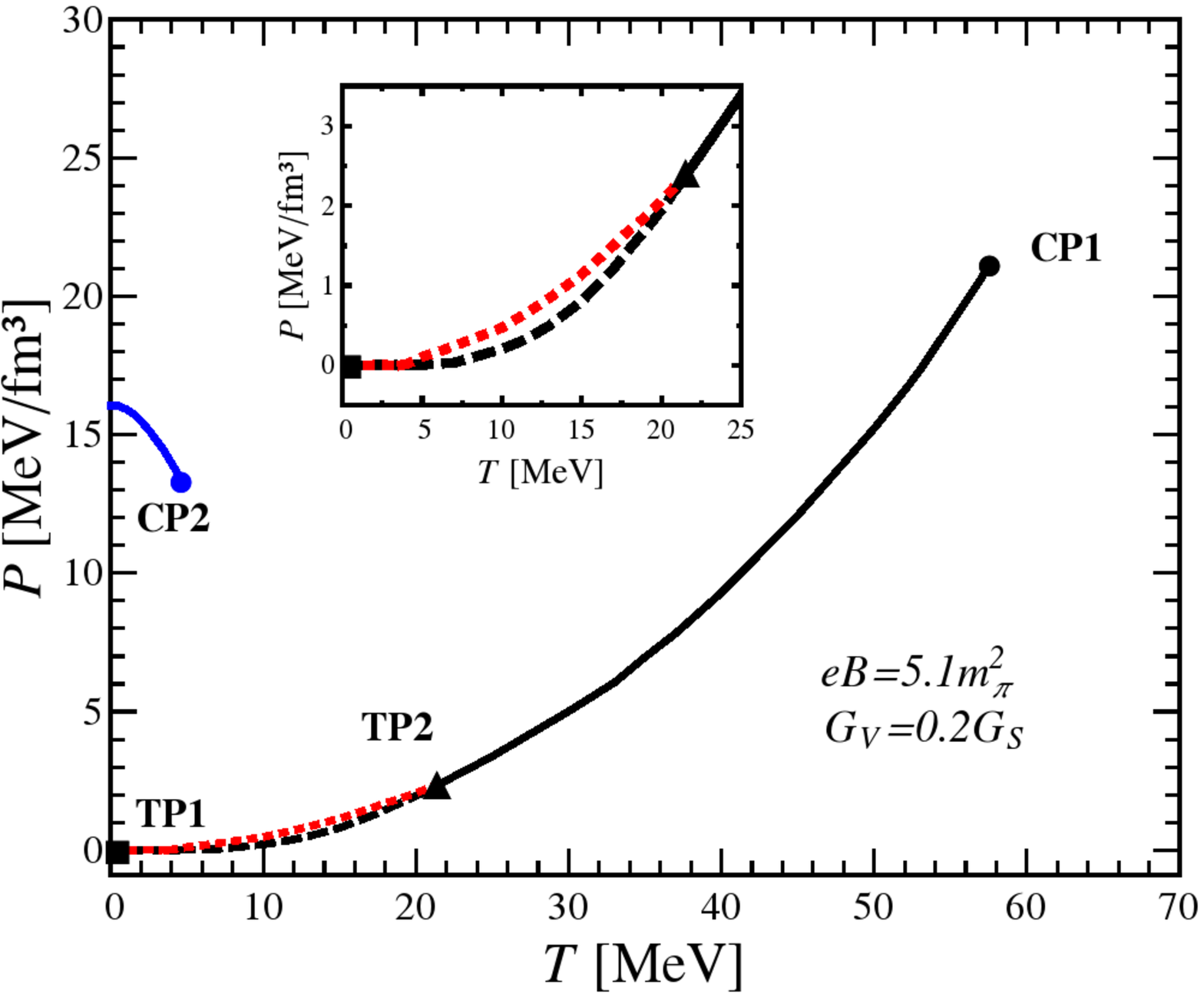} 
\caption{Phase diagrams in the $T-\mu$ plane (Left Panel) and in the $P-T$ plane (Right Panel) for$B=5.1\,m_{\pi}^2$ and  $G_V=0.2\,G_S$. At $T=0$,  both diagrams show a triple point (TP1) splits into two common 
dual-phase coexistence lines as the zoomed areas show. At $T_{TP2} = 21.4\,{\rm MeV}$, the dual coexistence lines converge again and merge at another triple point (TP2).
Beyond this temperature one observes the ordinary first order line  which ends in a critical point (CP1). The figure also shows that the  first order phase transition line which terminates at CP2 has a negative $\Delta P/\Delta T$ slope. }
\label{fig6}
\end{figure}

We are now in position to map all these transitions into  phase diagrams such as the ones shown in Fig. \ref {fig6} in order to  display the phase boundaries in  the $T-\mu$ and $P-T$ planes. 

\begin{figure}[htp]
\centering
\includegraphics[width=0.4\textwidth]{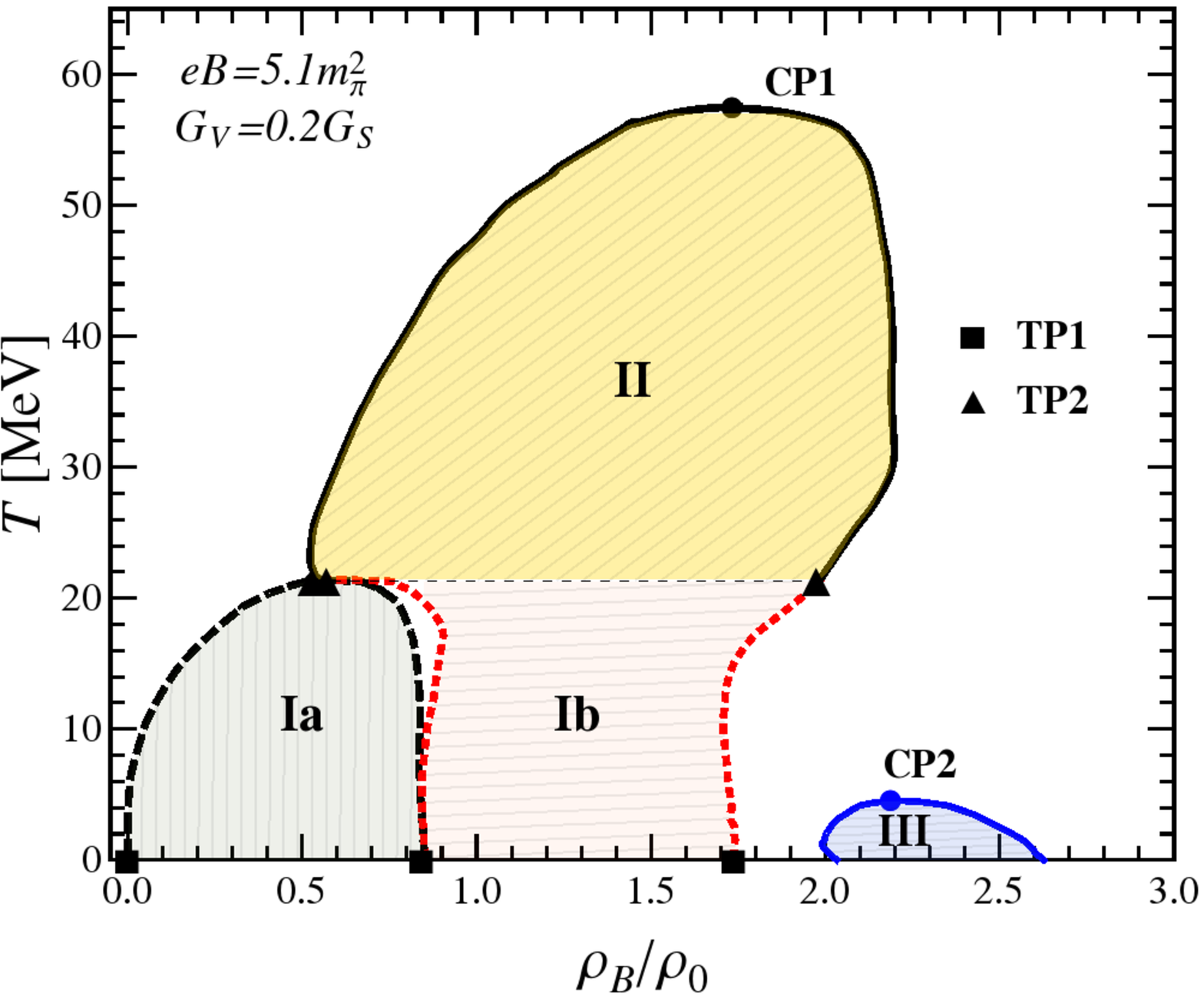} 
\hspace{10pt}
\includegraphics[width=0.4\textwidth]{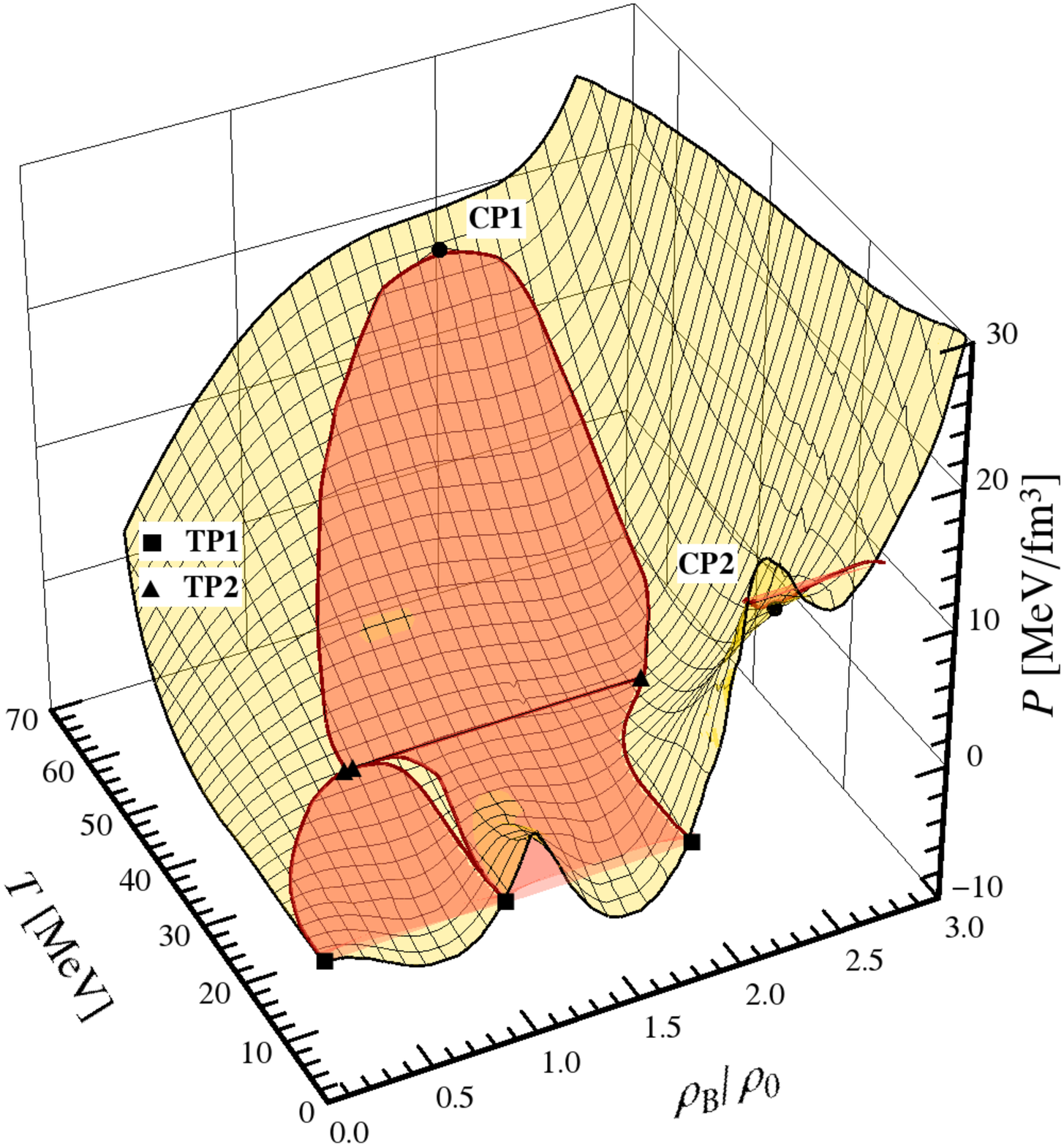} 
\caption{Phase diagram in the $T-\rho_B$ plane (Left Panel) and EOS isotherms in the $P-T-\rho_B$ space (Right Panel) for the $B=5.1\,m_{\pi}^2$ and $G_V=0.2\,G_S$ case.
Filled squares and triangles indicate the triple points and filled circles indicate the critical end points in accordance with the previous figures.}
\label{fig7}
\end{figure}

Analyzing these  figures we verify that the usual  phase diagrams are  drastically modified from the ordinary case at 
$G_V=0$ since we now have a first-order coexistence line originating from a triple point at ($T=0,\mu_{TP1}=388.05\,{\rm MeV}$) 
and then splitting into two new branches of ordinary first order transitions, as shown in the zoomed region, 
to finally merge   again at a second triple point  ($T_{TP2}=21.4 \,{\rm MeV},\mu_{TP2}=384.82 \,{\rm MeV}$). Then, for higher temperatures the 
transition always follows the usual pattern until the line ends a  critical point (CP1). The figure also 
shows an ordinary first order phase transition line which starting at ($T=0,\mu_{III}=404.70\,{\rm MeV}$) and terminating 
at CP2 ($T_{CP2}=4.75\, {\rm MeV},\,\mu_{CP2}=401.60 \,{\rm MeV}$). The $P-T$ allows to further understand these two different first order lines by 
observing the   $dP/dT$ slope in each case. The diagram shows that the line terminating at CP1 has a 
positive slope just like in the well known ``liquid-gas" type of transition while the line terminating 
at CP2 presents a negative slope which is reminiscent of the ``solid-liquid" transition occuring in the 
water phase diagram which does not appear to have been reported previously (at least within 
the context of strongly interacting systems).
The left panel of  Fig. \ref {fig7} displays the coexistence phase diagram in the $T-\rho_B$ plane while the right  panel of the same figure shows a three dimensional plot displaying the $P-T-\rho_B$ phase diagram with the Andrews isotherms.

It is well known that at zero temperature this type
of model, as well as the quark meson model,   typically
predict phase coexistence between the vacuum and the
dense quark phase, similar to the  ``liquid-gas"
transition. However, we now observe that when a vector repulsion and a magnetic field are  present it is 
possible that three phases coexist for particular parameter values. As Fig. \ref {fig7} shows, for $eB=5.1\,, m_\pi^2$ and $G_V=0.2 \, G_S$, 
one observes the coexistence between phases with densities $\rho_B/\rho_0= 0, 0.85\,, \rm{and} \,1.75$ which is a rather interesting result 
since the vanishing density represents the vacuum, the intermediate density value is very close to that of ordinary nuclear 
matter while the third is close to the values this model predicts for the dense quark phase. Eventually,  by scanning over 
parameter values one may force the intermediate density to take place at $\rho_B/\rho_0=1$ so that this could be identified 
as nuclear matter. As the temperature increases, the three-phase coexistence vanishes and one observes a decoupling into 
two separate branches $Ia$ and $Ib$, 
which are related to two first order transitions taking place at distinct pressure values. 
This bifurcation in two coexistence branches creates a stability island between the regions $Ia$ and $Ib$. 
Again, at $T_{TP2}=21.4\,{\rm MeV}$ these two distinct first order transitions will share the same
pressure at a triple point $TP2$ as the figures show. The remaining first order transition corresponding to the coexistence region $II$  ends at
an ordinary critical point $CP1$. Furthermore, as already emphasized, another coexistence region appears at a higher density range, 
labelled as $III$ in the left panel of Fig. \ref {fig7}. At $T=0$ the coxistence densities for this region are $\rho_B/\rho_0=2.0$ and $\rho_B/\rho_0=2.6$.
Then, when the temperature reaches the critical value  $T_{CP2}=4.75\, {\rm MeV}$ this coexistence region, which corresponds to 
a ``solid-liquid" type of transition, terminates at $CP2$. 
\par

\begin{figure}[htp]
\centering
\includegraphics[width=0.4\textwidth]{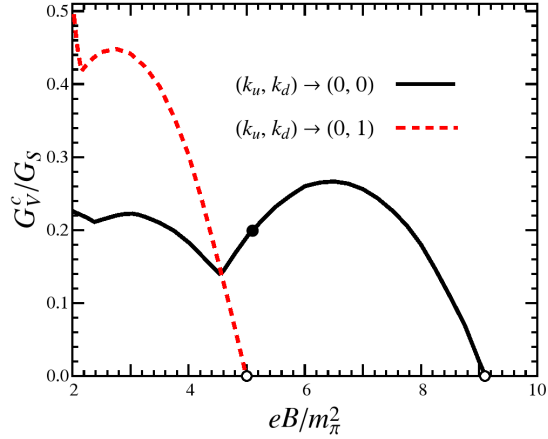} 

\caption{Magnetic field dependence of the critical values, $G_V^c$, related to the transitions $(k_u,\,k_d) \to (0,0)$  
and $(k_u,\,k_d) \to (0,1)$ 
 at $T=0$. The dot labels the values used in our analyzes and the open circles mark the excluded $G_V=0$ case.  }
\label{fig8}
\end{figure}

At this point it is important to recall that the occurrence of multiple phases at the same pressure is only possible if  thermodynamical potential develops multiple global minima. 
This condition can be attained to a particular choice of $B$ and $G_V$ values 
which give rise to multiple densities coexisting in the same transition chemical potential $\mu$. Let us now call $G^c_V$ the critical 
vector coupling value at which three degenerate global minima appear for a given $B$ value. Then, when mapping the parameter space of $(B,G^c_V,\mu)$ 
 values which result in a three-phase coexistence at $T=0$ we  encounter the oscillatory behavior shown  in  Fig. \ref{fig8}.
 
Each critical vector coupling $G^c_V$ can be associated with the emergence of a new   intermediary (stable) 
value of the order parameter. Depending on the field strength, more than one intermediary Landau level configuration $(k_u,\,k_d)$ 
can appear and each one has its own $G^c_V$ value. 
So, it must exist an infinite set
of critical values $G^c_V$  associated with the infinite number of intermediary states $(k_u,\,k_d)$ in the limit of $B \to 0$ and 
$G_V \to \infty $. The figure  only shows the first two possibilities characterized by the transitions
 $(k_u,\,k_d) \to (0,0)$  and $(k_u,\,k_d) \to (0,1)$. As one can observe the former transition covers the region of  the parameter values which are usually considered in the recent literature  ($G_V = 0-0.5\,G_S$ and $eB = 2-8\,m_\pi^2$). Note that for high fields $eB \gtrsim 9\,m_\pi^2$ the emergence of triple points is suppressed since only the LLL is occupied. We have tested coupling values above and 
 below the critical line observing that when $G_V < G^c_V$  only the traditional case with one first order transition 
 line emerges. Then,  when $G_V = G^c_V$ the multiple coexistence  becomes possible and a new first order transition 
 lines starts from the original one at $T=0$ as shown in Figs. \ref{fig6} and \ref{fig7}.
Finally, when $G_V > G^c_V$ the chiral symmetry restoration takes place via more than one first order transition line in a cascade mode as discussed in Section III. We close this section by remarking that we have also observed the coexistence of four phases but this is resctrited to the $T=0$ case only.
 

\section{ Conclusions}
We have investigated how the presence of a magnetic field and a repulsive vector interaction may influence the phase diagram of strongly interacting matter generating unusual transition patterns associated to first order chiral transitions. In the first part of the present study we have considered parameter values which produce a cascade of first order transitions similar to the ones analyzed in Refs. \cite {ebert, Allen:2013lda, Grunfeld:2014qfa} at vanishing temperatures. We have taken a step forward by incorporating the vector interaction as well as by pushing the evaluations to the finite temperature domain in order to better understand the physical nature of such transitions. Mapping the transition into the $P-T$ plane has allowed us to observe  that the  transition which takes place at high pressure  has a negative $dP/dT$ slope which is reminiscent of the ``solid-liquid" transition observed in the water phase diagram while the remaining (low pressure) transitions have a positive slope and therefore resemble the  ``liquid-gas" transition which is  usually observed within effective quark models. Having in mind that, due to the inherent  oscillations  caused by the filling of Landau levels, the magnetic field induces the free energy to develop multiple minima while the repulsive vector interaction favors stability we have scanned over $B$ and $G_V$ values to check for the possibility of observing three degenerate global minima which would then lead to the existence of a triple point. Our expectation was confirmed by the numerical investigation and we were able to find, for a given value of $B$, a certain value of $G_V$ so that three (instead of the usual two) phases coexist which, as far as we know, has not been observed before. In contrast to the ``solid-liquid" type of transition, which may be find when canonical parametrizations are used, the existence of triple points is only possible when very specific values of $G_V$ are chosen. Finally, we point out  remark that the coexistence of multiple phases has also been recently observed by Manso and Ramos \cite {rudnei} within the $2+1\;d$ Gross-Neveu model despite the fact that  a repulsive term has not been considered. Instead the authors have considered a tilted magnetic field observing that the parallel component tends to stabilize the free energy (just as $G_V$ in our case) so that three degenerate minima were also observed.  Together with our observations this result leads us to conclude that the phase diagram of magnetized matter may display the coexistence of multiple phases if the dynamics contains terms which bring stability to the free energy. 

\section*{Acknowledgments}

This work was  partially supported by Conselho Nacional de Desenvolvimento Cient\'{\i}fico e Tecnol\'{o}gico (CNPq).

\end{document}